\documentclass[12pt]{article}
\usepackage{amsmath}
\usepackage{amssymb}
\usepackage{latexsym}
\usepackage{epsfig}

\newcommand{\be}{\begin{equation}}
\newcommand{\ee}{\end{equation}}

\newcommand{\sectiono}[1]{\section{#1}\setcounter{equation}{0}}

\newcommand{\s}{ {\bf s } }
\renewcommand{\t}{  {\bf t} }
\renewcommand{\u}{ {\bf u } }

\begin{document}

{}~ \hfill\vbox{
\hbox{CTP-MIT-3856}
\hbox{YITP-SB-07-27}} \break

\vskip 2.0cm
\centerline{\large \bf The off-shell Veneziano amplitude in Schnabl gauge}
\vskip 0.3cm

\vspace*{9.0ex}
\centerline{ \sc Leonardo Rastelli}

\vspace*{2.5ex}

\centerline{ \it  C.N. Yang  Institute for Theoretical Physics}
\centerline{ \it Stony Brook University}
\centerline{ \it  Stony Brook, NY 11794, USA}

\vspace*{3.5ex}

\centerline{ \sc Barton Zwiebach}

\vspace*{2.5ex}

\centerline{ \it Center for Theoretical Physics}
\centerline{ \it Massachussetts Institute of Technology}
\centerline{ \it  Cambridge, MA 02139, USA}

\vspace*{6.5ex}
\medskip
\centerline {\bf Abstract }

\bigskip
\bigskip

We give a careful definition of the open string propagator in Schnabl gauge
and present its worldsheet interpretation.  The propagator requires two Schwinger parameters and contains the BRST operator.  It builds surfaces 
by gluing strips of variable width to the left and to the right of off-shell states
with contracted or expanded local frames.
We evaluate explicitly  the  four-point amplitude of off-shell tachyons.
The computation involves a subtle boundary term, crucial
to enforce  the correct exchange symmetries. 
Interestingly, the familiar on-shell physics emerges even though
string diagrams produce Riemann surfaces more than once.  Off-shell,
the amplitudes do not factorize over intermediate on-shell states.

\bigskip

\vfill \eject

\baselineskip=18pt

\tableofcontents  

\sectiono{Introduction} 

Off-shell amplitudes have been the subject of 
much interest
throughout the history of string theory. 
It was suspected from the beginning that the celebrated Veneziano 
four-point amplitude~\cite{Veneziano:1968yb}
for the on-shell scattering of  open string tachyons would have a
sensible off-shell extension.  Even three-point vertices would have
off-shell extensions. It was clear that a consistent {\em set} of off-shell
amplitudes would emerge from a field theory of strings.

Although the amplitudes of light-cone string field theory~\cite{Kaku:1974zz} make sense off-shell,
their properties are unusual. With the development of  a Lorentz covariant open string field
theory~\cite{Witten:1985cc}, the study of  off-shell amplitudes began in earnest. 
These studies  had useful applications.  In fact, 
the expected properties of off-shell closed string amplitudes  
gave contraints~\cite{Sonoda:1989wa,Sonoda:1989sj}  
that helped in the construction of closed string field theory \cite{Zwiebach:1992ie}.
Off-shell amplitudes were studied mostly using the Siegel gauge. 
It was learned that:

\begin{enumerate}

\item  The amplitudes have permutation symmetry
among scattering states.

\item The amplitudes are   
integrals over sections of fiber bundles
with base the moduli spaces of Riemann surfaces and fibers 
spanning the possible choices of local coordinates at the 
punctures where the scattering states are inserted. 
Ignoring coordinates at the punctures, each Riemann surface
contributes only once to the amplitude.

\item  The amplitudes satisfy {\em  factorization}:
 near 
 poles, all of which must arise from the propagator, the amplitude
 is a product of the relevant off-shell vertices. 
 
\end{enumerate}

The first property arises because the vertices in the string field
theory action are symmetric and so is the 
propagator.\footnote{In open string field theory the 
vertices are only cyclically symmetric, full symmetry arises
by summing over inequivalent orderings.  In closed string field theory the vertices
are, by construction, fully symmetric.}  The second property implies that the
string diagrams for a given amplitude give a construction of the 
appropriate moduli space
of Riemann surfaces:  they produce all surfaces of fixed genus and fixed number 
of punctures, each surfaced produced only once.  The third property arises
because string diagrams at factorization develop infinitely long strips
(or tubes, in closed string theory) that if cut, result in  two allowed
subdiagrams\footnote{There are also factorizations in which cutting a line does 
not split the diagram into two separate pieces.} 
which provide the two off-shell factors.

Siegel gauge provides amplitudes that obey the above
properties,   but not all gauges do.  Off-shell light-cone amplitudes
do not satisfy property 1 because the string diagrams break the symmetry 
among states by assigning to them values of the light-cone momentum $p^+$
all of which cannot be the same.  They do not satisfy property 3 either,
because the Schwinger parameters associated with propagators are sometimes not independent.

In this paper we begin a detailed study of off-shell amplitudes in Schnabl gauge,
the gauge in which it was possible to obtain an analytic form
for the tachyon vacuum string field~\cite{Schnabl:2005gv}.
The string field that represents a finite marginal deformation by a regular marginal
operator is another solution  in Schnabl gauge~\cite{Schnabl:2007az,KORZ}.
Other analytic solutions~\cite{Fuchs:2007gw,Kiermaier:2007vu,Erler:2007rh,Kishimoto:2007bb} use the
wedge states \cite{RZ, RSZ1, RSZ2, martinwedges, bonora} that  are natural in Schnabl gauge, but
do not actually satisfy the  gauge condition. 
Recent related  work  appears in \cite{yuji}--\cite{relatedlast}.

The simplest amplitude to consider is the Veneziano amplitude.  Its off-shell
version in Schnabl gauge is the central topic in this paper.  In Siegel gauge
the Veneziano amplitude was first discussed by Giddings~\cite{Giddings:1986iy}, 
who found 
 the conformal map from the string diagram to the upper-half plane and then
showed that the amplitude would reproduce the familiar on-shell result.
The closed-form expression for the amplitude is fairly complicated as it
requires constraints that involve elliptic functions and their inverses. 
Further analysis and applications were discussed
 in~\cite{Giddings:1986bp,Samuel:1987uu}.  
 
The off-shell Veneziano amplitude in Schnabl gauge 
was first examined in the useful
paper by Fuji, Nakayama, and Suzuki~\cite{japanese}, who used an
algebraic approach. 
We use the more geometrical conformal field theory
interpretation of the amplitude and emphasize at each step the 
integrations over the moduli of the relevant Riemann surfaces. As it turns
out, we have found that a subtle boundary term is 
missing\footnote{There are some further small disagreements
in our computations.} 
in the computation of \cite{japanese}.   This term is needed because the naive computation
gives a result that violates the manifest exchange symmetry between incoming
and outgoing states of the string amplitude.

As first stated in~\cite{Schnabl:2005gv}, and  
 as we review in \S\ref{thepropagator}, a propagator
$\mathcal{P}$ that {\em formally} inverts the BRST operator
 in the gauge $B \Psi = 0$ is  
\be
\label{schnabl_gauge_propagator}
\mathcal{P} \equiv \frac{B}{L} \,Q\, \frac{B^\star}{L^\star} \, .
\ee
Here $\star$ denotes BPZ conjugation, $Q$ is the BRST operator, and $B$ and $L$ denote, respectively,
the antighost  and the Virasoro zero modes in the sliver frame.\footnote{Our conventions
are the same as in \cite{RZ, ORZ, KORZ} and are reviewed in appendix A.}  In Siegel gauge
each propagator $b_0/L_0$ uses one Schwinger parameter, exactly 
the number needed to produce the moduli space of surfaces with punctures on the boundary.
In Schnabl gauge the propagator uses two Schwinger parameters, one  to represent $1/L$ and another
to represent $1/L^\star$.  With two parameters for each internal line 
we overcount moduli space and each surface
is produced an infinite number of times. We exhibit this phenomenon 
explicitly
for the off-shell Veneziano amplitude.  It is the presence of $Q$ that makes this overcounting
compatible with the familiar on-shell result.

In computations one may  use 
$\{Q, B^\star\} = L^\star$ and naively
assume that this  $L^\star$ 
cancels with the $1/L^\star$ in $\mathcal{P}$ to give 
\be
\label{schnabl_gauge_propagator_naive}
\mathcal{P} ~\overset{?}{=} ~\frac{B}{L} -  \frac{B}{L}\frac{B^\star}{L^\star}  \,Q\,\, .
\ee
This is the propagator used in~\cite{japanese} to compute the 
Veneziano amplitude\footnote{The presentation of B. Zwiebach in Strings 2007 (Madrid)
also used this propagator.}.  By looking at the factorization properties of the $s$-channel
contribution to the off-shell four-point amplitude following from 
(\ref{schnabl_gauge_propagator_naive}) we find that the exchange symmetry
under $(p_1, p_2) \leftrightarrow (p_3, p_4)$ is violated.  
We then demonstrate that the requisite symmetry is restored if we include
on the right-hand side of  (\ref{schnabl_gauge_propagator_naive}) an extra
boundary term that arises when $1/L^\star$ is represented by a Schwinger parameter
with a cutoff $\Lambda^\star$:
\be
\label{btersm}
- \lim_{\Lambda^\star \to \infty} \,{1\over 2}\,\frac{B}{L} e^{-\Lambda^\star L^\star} \,.
\ee
We explain  how this term arises from a regulated version of 
(\ref{schnabl_gauge_propagator}) where the  Schwinger
representation of the propagator uses two independent cutoffs $\Lambda$ and 
$\Lambda^\star$ and a symmetric limit is used to take them to infinity.  
We check that this concrete  definition for the propagator provides 
an inverse for the BRST operator. 
Further analysis shows that the boundary term that arises  
from (\ref{btersm}) is actually
antisymmetric under the exchange.  Hence
a symmetrized version of the propagator (where $Q$ is moved both to the left
and to the right) requires no
boundary terms.
We conclude that, after proper care of all subtleties,
property 1 holds. 

Property 2 does not hold.   In Schnabl gauge off-shell amplitudes include 
 finite integrals over
parameters that {\em do not} change the moduli; 
the string diagrams  produce Riemann surfaces  more than once. 
This is illustrated 
in the Veneziano amplitude, where we find a coordinate
for the redundant direction of integration that simplifies the amplitude considerably.
The possible existence of a set of consistent off-shell amplitudes that
are not built  
by integration over moduli space is probably the most
novel and nontrivial feature of the Schnabl gauge.  

We were somewhat surprised that property 3 does not hold --
  the off-shell Veneziano amplitude does not exhibit factorization.  
  This is checked, at lowest level, by looking at the pole that arises when the intermediate
line is an on-shell tachyon.  We find that the residue is {\em not}
the product of two Witten vertices with two lines off-shell and one line on-shell. 
It is, instead, 
 the symmetrized product of two different vertices.  
Of course, the string diagrams do not suggest off-shell 
factorization since there is no long strip separating identical looking vertices. 

\smallskip
This paper begins in \S\ref{thepropagator} with an explanation
for (\ref{schnabl_gauge_propagator}).  We then discuss the CFT interpretation
of $B/L$ acting on star products of off-shell states.  Our result here is a generalization
of the on-shell result in~\cite{KORZ}.  We find
a related CFT interpretation for the action of $B^\star/L^*$.  These formulas
are the analogs of the simple strip-plus-antighost
representation of the Siegel gauge propagator $b_0/L_0$.  In
\S3 we 
compute the two diagrams that correspond 
(\ref{schnabl_gauge_propagator_naive}).  
We rewrite
the amplitudes in terms of moduli in order to simplify the results.  
As would be expected,
the formulas  are much 
simpler to write and evaluate than the corresponding ones in Siegel gauge.
By examination of the factorization on on-shell tachyons, we note that the
amplitude computed so far fails to have the
expected  symmetry under the exchange 
$(p_1, p_2) \leftrightarrow (p_3, p_4)$.  In \S4  we identity 
  the culprit in a subtle
boundary term that happens not to vanish.
We explain how the boundary term arises
from a suitably regulated  propagator 
 and demonstrate  that its addition restores the exchange symmetry. 
 We revisit the formal arguments of \S\ref{derprop} and show that
 with this prescription the propagator provides an inverse for the
 BRST operator.  
 The final formulae for the off-shell Veneziano amplitude are 
 given in (\ref{Final}) and (\ref{osvam}).   In \S5 we discuss
our results and some interesting open questions.  A collection of useful
formulas have been relegated to three appendices.

\sectiono{The propagator, $B/L$, and $B^\star/L^\star$}\label{thepropagator}

\subsection{Deriving the propagator}\label{derprop}

 An open string field $\Psi$ is in  Schnabl gauge if it satisfies
\be
\label{sch_gauge_cond}
B \Psi = 0 \, .
\ee
Here $B$ denotes the antighost zero mode in the conformal
frame of the sliver and is a linear combination   
 of $b_{2n}$ oscillators, $n \geq 0$.
The operator $B$ is not BPZ even,  $B^\star \not= B$.  Clearly, the operator
$B$ squares to zero, $B^2=0$. 
The full state space $\mathcal{H}$ breaks into two complementary vector subspaces $S$ and $S'$,
where $S$ is the subspace of  states that satisfy the gauge condition
(\ref{sch_gauge_cond}).  To show 
this consider an operator $C$ such that $C^2 = 0$ and $ \{ B, C\} = 1$.
The form of $C$ is not relevant, but one could take $C=c_0$, for example.  
While $c_0^\star = - c_0$, in general $C^\star \not= - C$.  From BPZ conjugation
rules it follows that $\{ B^\star , C^\star\} = -1\,.$
Then introduce the orthogonal projectors 
\be
P= B C\,, \quad P' = C B\,.
\ee
Indeed, one readily verifies that $P^2=P, ~P'^2 = P', ~PP'= P'P=0$,  and $P + P'=1$.
We have $P: \mathcal{H}\to S$ because $B P =0$.   We also note that
\be
P^\star =  - C^\star B^\star\,.
\ee
In  Siegel gauge ($b_0\Psi =0$) 
one has $P = b_0 c_0$,  $P'= c_0b_0$, and $P^\star  
= - c_0^\star b_0^\star =  c_0 b_0 = P'$.
In Schnabl gauge, $P^\star \neq P'$. 

To find the propagator one considers the kinetic operator 
$\cal{K}$ obtained by restricting the kinetic term to string fields in the gauge slice, {\it i.e.} string fields of the form $P\Psi$:
\be
{1\over 2} \langle P\Psi, Q P\Psi\rangle = {1\over 2} \langle \Psi, P^\star Q P\Psi\rangle 
\equiv  {1\over 2} \langle \Psi, \mathcal{K} \Psi\rangle \,, \qquad
\mathcal{K} =  P^\star Q P \,. 
\ee
In Siegel gauge
${\cal K} = c_0 b_0 Q b_0 c_0 = c_0 L_0 b_0 c_0  = c_0 L_0$.  Assuming $L_0$ has
an inverse $1/L_0$ one can almost invert ${\cal K}$, except for the fact that $c_0$ is
not invertible.  One introduces a propagator $\mathcal{P}$ such that
$\mathcal{P} \mathcal{K}$ equals the projector to the gauge slice.  This is easily
done
\be
\mathcal{P} ={b_0\over L_0} \quad \hbox{gives} \quad \mathcal{P} {\cal K} 
= {b_0\over L_0} c_0 L_0 = b_0 c_0 = P\,.
\ee
This is reasonable:  when solving ${\cal K} \Psi = J$, the action of $\mathcal{P}$
on both sides gives  
$P\Psi = \mathcal{P}J$ and both left- and right-hand sides are clearly in the gauge.

In  Schnabl gauge the kinetic term is 
\be
{\cal K} = P^\star Q P = - C^\star B^\star Q B C\,.
\ee
As opposed to the situation in Siegel gauge, this ${\cal K}$ cannot be simplified. 
The propagator $\mathcal{P}$ must give  $\mathcal{P} {\cal K} = P$, namely
\be
-\mathcal{P}\,C^\star B^\star Q B C  =  BC\,.
\ee
We note that $\mathcal{P}$ cannot be equal to $B/L$.  
We must have $\mathcal{P} = B \ldots B^\star$, the left-most $B$ is there to leave us
in the gauge ($B\mathcal{P} =0$) and the right-most $B^\star$ to cancel the
$C^\star$ in ${\cal K}$. 
As first noticed in \cite{Schnabl:2005gv},  the propagator takes
the form
\be
\label{claimed}
\mathcal{P} \equiv \frac{B}{L} \,Q\, \frac{B^\star}{L^\star} \, .
\ee
Indeed, as desired
\be
\mathcal{P} {\cal K} = -\frac{B}{L} \,Q \,\frac{B^\star}{L^\star}\,C^\star B^\star Q B C =
\frac{B}{L}\, {Q\over L^\star}  B^\star Q B C =
\frac{B}{L}  Q B C = BC\,.
\ee
One can similarly check that ${\cal K} \mathcal{P} = P^\star$.  
It is very important to note that the propagator $\mathcal{P}$ is BPZ even:
\be
\begin{split}
\langle{R_{12}} | \mathcal{P}^{(1)} = \langle R_{12} |\frac{B^{(1)}}{L^{(1)}} \,Q^{(1)}\, \frac{{B^{(1)}}^\star}{{L^{(1)}}^\star}
 &= \langle R_{12} | \,Q^{(1)}\, \frac{{B^{(1)}}^\star}{{L^{(1)}}^\star}
 \frac{{B^{(2)}}^\star}{{L^{(2)}}^\star} \\
 &= \langle R_{12} | \,(-Q^{(2)})\, \frac{{B^{(1)}}^\star}{{L^{(1)}}^\star}
 \frac{{B^{(2)}}^\star}{{L^{(2)}}^\star}\\
  &= \langle R_{12} | \frac{{B^{(2)}}}{{L^{(2)}}}\,Q^{(2)}\,
 \frac{{B^{(2)}}^\star}{{L^{(2)}}^\star}  = \langle{R_{12}} | \mathcal{P}^{(2)} \,.
 \end{split}
\ee 
We simply write $\mathcal{P}^\star = \mathcal{P}$.  

The manipulations leading to the derivation of $\mathcal{P}$ have
been somewhat formal --   
we cancel $L$'s against
$1/L$'s with impunity.  As we will see later, there are subtleties that invalidate
such cancellations unless a suitably regulated definition of the propagator
is used.

\subsection{CFT representation of $B/L$}

Since $B/L$ and $B^\star /L^*$ enter in the propagator
(\ref{claimed}),  we must understand how these operators act on states and
their star products.  The CFT representation of this action will be used to
obtain the string diagrams associated with this gauge.
In this subsection we find the CFT representation
of the operator $B/L$.  In the following subsection with study the action
of $B^\star/L^\star$.

Let us begin by considering the action of $1/L$.  For arbitrary
Fock space states $A_1$ and $A_2$ we have
\begin{equation}
\label{1/L_on_two_product}
\begin{split}
\frac{1}{L}  ( A_1 * A_2 )
 = & \int_0^\infty dT \,  e^{-T L}  (A_1 * A_2 )  
 =  \int_0^\infty dT  \,  e^{-T L} A_1 *  e^{-T (L- L^+_L)} 
 A_2  \,, \\ 
 = & \int_0^\infty dT \,  e^{-T L} A_1 *  
 e^{ (1-e^{-T}) L^+_L}e^{-T L} A_2\,, \\  
  = & \int_0^1 {dt\over t}  \, \,  t^L A_1 *  
W_{t-1} * t^L  A_2  \,.
\end{split}
\end{equation}
Here we have used (\ref{large_act}), (\ref{ordering1}), and  
$e^{-\alpha L^+_L} A = W_\alpha *A $, which holds for arbitrary $A$.  
For multiple string fields this generalizes to
\begin{equation}
\label{1/L_on_prod}
 \frac{1}{L}  ( A_1 * A_2 *  \ldots  A_k )
 =  \int_0^1 {dt\over t}  \, \,  t^L A_1\, * \,
(W_{t-1} *t^L  A_2)  *
\ldots  *  \,
(W_{t-1} * t^L  A_k ) \,.
\end{equation}
There is a wedge $W_{t-1}$  between every consecutive pair of states
 $t^L A_i$ and $t^L A_{i+1}$.  Given that $0\leq t \leq 1$, this wedge
 state removes a piece of surface.
 
\begin{figure}
\centerline{\hbox{\epsfig{figure=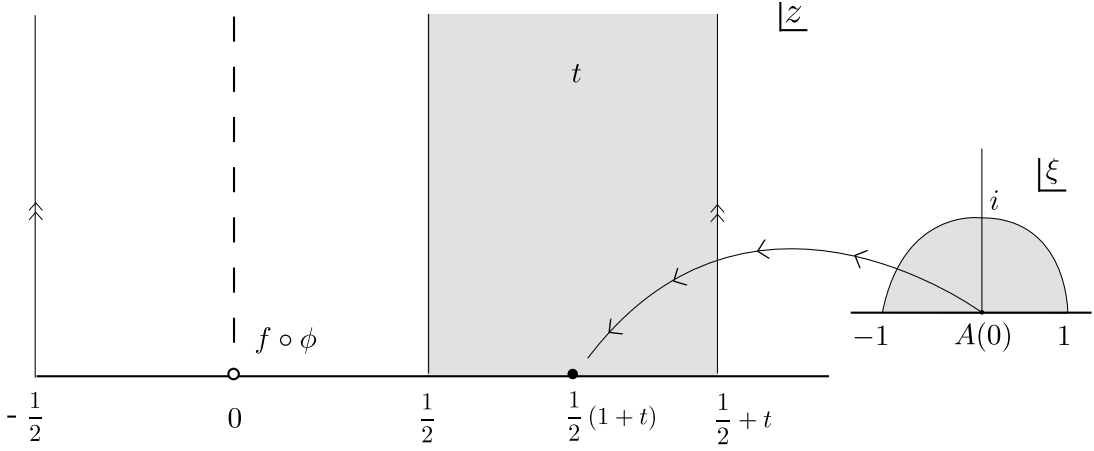, height=5.9cm}}}
\caption{The geometrical picture for the state $[A]_t$, a wedge state $W_t$ with
$A$  inserted using a local coordinate that covers the width $t$ strip.}
\label{rz1fig}
\end{figure}

 It will be convenient to introduce the state $[A]_t$ associated to the state $A$
 and defined by
\be
\label{bracket_notation}
[A]_t \equiv W_{{1\over 2} (t-1)} * t^L A * W_{{1\over 2} (t-1)} \,. 
\ee
The state $[A]_t$ has an interesting geometrical picture, shown in Figure 1.
The state $t^L A$ can be visualized as
the unit wedge $-{1\over 2} \leq \hbox{Re}(z) \leq {1\over 2}, ~
\hbox{Im}(z) \geq 0$,  with the operator
$t^L A(\xi=0)$ mapped to $z=0$ by means~of 
\be
z= f(\xi) = {2\over \pi} \tan^{-1} \xi\,.
\ee
The state $[A]_t$ is obtained  by 
gluing a wedge of width ${1\over 2} (t-1)$ to the unit-width $t^LA$ wedge,
 followed by the gluing of another wedge of width ${1\over 2} (t-1)$.
All together we have a wedge of width $2\cdot {1\over 2} (t-1) + 1 = t$.  In the CFT
language the test state  is inserted on the canonical unit wedge and the wedge
$[A]_t$ is built starting at $z= {1\over 2}$. The local insertion then occurs at
${1\over 2} + {1\over 2} (t-1) + {1\over 2} = {1\over 2} (1+t)$, so we have
\be
\langle \phi\,, [A]_t\rangle =  \bigl\langle f\circ \phi(0) \,  f_{{1\over 2} (1+ t)} \circ t^L A (0)
\bigr\rangle_{W_t}\,,\ee
where
\be
f_r (\xi) =  r + f(\xi) =  T_r \circ f (\xi)\,, \quad  \hbox{with}\quad T_r (z) = z + r\,.
\ee
Indeed, as constructed,  the state $t^LA$ lands at $z = {1\over 2} (1+ t)$ 
(see Figure~\ref{rz1fig}).  The effect of $t^L$ on the state $A$ is that of a 
conformal map, so we can determine what is the full conformal map applied
to the state $A$.  As shown in~\cite{RZ} 
\be
t^L  \,\phi (\xi)\, t^{-L}  = ( f^{-1} \circ t \circ f) \circ \phi(\xi)\,,  
\ee
where $t$ denotes the map  $ z\to t z$.  It follows that
\be
f_{{1\over 2} (1+ t)} \circ t^L A (0) = T_{{1\over 2} (1+ t)} \circ f \circ f^{-1} \circ t \circ f
\circ A =  T_{{1\over 2} (1+ t)}  \circ t \circ f \circ A \,.
\ee
As we can see, the effect of $t^L$ is just to scale the local coordinate by a factor  of $t$.  The local coordinate domain for the $A$ insertion is  thus a wedge
of width $t$.  Since we began with a wedge of unit width, the extra wedge
state factors to the left and to the right, provide precisely the missing piece of surface.
All in all,
\be
\boxed{ 
\begin{split}
\phantom{\Bigl(}  
&[A]_t~\hbox{ is a wedge state of width}~ t ~
\hbox{ with} ~A~\hbox{ inserted at the}~ \\
&\hbox{boundary midpoint and the local patch filling the wedge.}
\phantom{\Bigl(}
\end{split}}
\ee

It follows from (\ref{bracket_notation}) that 
\be
t^L A   =  \, W_{{1\over 2} (1-t)} *  [A]_t * W_{{1\over 2} (1-t)} \, . 
\ee
We can use this to rewrite
(\ref{1/L_on_prod}) as
\begin{equation}
\label{1/L_on_prod_ov}
\frac{1}{L}  ( A_1 * A_2 *  \ldots  A_k )
 =  \int_0^1 {dt\over t}  \, \, W_{{1\over 2} (1-t)}* [A_1]_t\, * \,
[ A_2]_t  *
\ldots  *  \,
[A_k]_t *  W_{{1\over 2} (1-t)}\,.
\end{equation}
The geometrical picture is that of a sequence of glued wedges of width $t$, each
with a local insertion, flanked from the left and from the right with wedges of 
size ${1\over 2} (1-t)$.

\bigskip
\noindent
Now we proceed with the action of $B$, {\em without} assuming that
the states are in Schnabl gauge.  We do this on the product of two states,
beginning with the result in (\ref{1/L_on_two_product}). 
The structure of terms requires the
evaluation of 
\be
M \equiv  (B-B^+_L)
W_{t-1} * t^L  A_2 ~+ ~ W_{t-1} *  (B-B^+_L) t^L  A_2 \,.
\ee
Using $(B - B^+_L) W_\alpha = -\alpha B^+_L W_\alpha$, we get
\be
M=  - tB^+_L
W_{t-1} * t^L  A_2 ~+ ~W_{t-1} *   t^L  BA_2 \,.
\ee
The calculation is then straightforward 
\begin{equation}
\begin{split}
\frac{B}{L}  ( A_1 * A_2 )
 =&  \int_0^1 {dt\over t}  \, \Bigl[ ~~ t^L BA_1 *  
W_{t-1} * t^L  A_2 \\
&\qquad\qquad   + (-1)^{A_1+1} t^{L+1} A_1 *  B^+_L
W_{t-1} * t^L  A_2   \\
& \qquad\qquad  + (-1)^{A_1} t^L A_1 * 
W_{t-1} *   t^L  BA_2   \Bigr]   \,.
\end{split}
\end{equation}
Rearranging we get
\begin{equation}
\begin{split}
\frac{B}{L}  ( A_1 * A_2 )
=&~ (-1)^{A_1+1}\int_0^1 dt\,~ t^{L} A_1 *  B^+_L
W_{t-1} * t^L  A_2   \\
 +&  \int_0^1 {dt\over t}  \, \Bigl[~ t^L BA_1 *  
W_{t-1} * t^L  A_2  + (-1)^{A_1} t^L A_1 * 
W_{t-1} *   t^L  BA_2 \Bigr]  ~ \,. 
\end{split}
\end{equation}  
\begin{figure}
\centerline{\hbox{\epsfig{figure=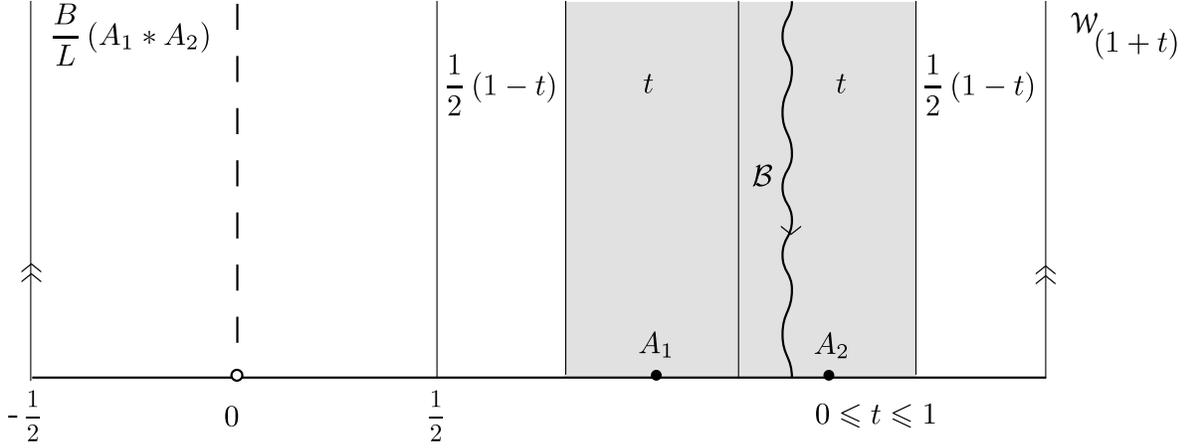, height=5.9cm}}}
\caption{A wedge state in the integral representation 
(\ref{the_B?Laction}) of ${B\over L} (A_1 *A_2)$.
Shown is one term only, the other two terms  have no  
$\mathcal{B}$ insertion but rather local insertions of $BA_1$ or $BA_2$. Note the
wedges of width ${1\over 2} (1-t)$, with $t\in [0,1]$, to the sides of $[A_1]_t*[A_2]_t$.}
\label{rz2fig}
\end{figure}
In terms of states of type $[\ldots ]$, 
\begin{equation}
\label{the_B?Laction}
\boxed{
\begin{split} \phantom{\Biggl(}
\frac{B}{L}  ( A_1 * A_2 )
=&~ (-1)^{A_1+1}\int_0^1 dt\,~ W_{{1\over 2} (1-t)} *  [A]_t *  B^+_L
 [A_2]_t * W_{{1\over 2} (1-t)}  \\
 & \hskip-20pt + \int_0^1 {dt\over t}  \,W_{{1\over 2} (1-t)} *\Bigl[~ [BA_1]_t *  
[A_2]_t  + (-1)^{A_1}[A_1]_t * 
[BA_2]_t \Bigr]* W_{{1\over 2} (1-t)}  \,.\phantom{\Biggr)}
\end{split}}
\end{equation}
The first line on the right-hand side of (\ref{the_B?Laction})
 is represented in Figure~\ref{rz2fig}.
This expression makes it manifest that
the  local coordinate patches of $A_1$ and $A_2$ match seamlessly.
Moreover, extra strips of width ${1\over 2} (1-t)$ are added to the left and to the right
of the $[A]_t$ and $[A]_2$ wedges, as shown in the figure. 
For one state only we have
\be
{B\over L} A = \int_0^1 {dt\over t} \,W_{{1\over 2} (1-t)} *  [BA]_t 
* W_{{1\over 2}(1-t) } \,.
\ee

\subsection{CFT representation of $B^\star/L^\star$} 

We begin with the evaluation of $1/L^\star$ acting on a single
state:
\be
{1\over L^\star} A = \int_0^\infty \, dT\,e^{-TL^\star}  A = \int_0^\infty \,dT\,
e^{-T(L^+ - L)} A  \,.
\ee
The exponential can be broken up using (\ref{cbh_2_dim}): 
\be
{1\over L^*} A = \int_0^\infty \,dT\,
e^{(1-e^T) L^+}  e^{TL} A   = \int_1^\infty  {ds\over s}  \, e^{(1-s)L^+} \, s^L A
\ee  
\be
{1\over L^\star} A = \int_1^\infty \hskip-2pt  {ds\over s}  \,\, W_{s-1} *\, s^L A* W_{s-1}\,. 
\ee
To generalize we recall the second identity in (\ref{large_act}),
and noting that $L^\star - L^+_L = L^+_R-L$, we get
\be 
e^{- T L^\star }  ( \phi_1 * \phi_2  * \cdots \phi_n ) = e^{-T L^\star} \phi_1  *
 e^{T (L-L^+_R)} \phi_2  * \cdots    e^{T (L-L^+_R)} \phi_n \,.
\ee
With this and (\ref{ordering2}) 
we find
\begin{eqnarray} 
{1\over L^\star} (A_1 * A_2 * \ldots *A_k) & 
=  & \int_0^\infty dT \, e^{-TL^\star} A_1 * e^{T (L-L^+_R)} A_2 * \ldots 
* e^{T (L-L^+_R)}A_k\\
&= & \int_0^\infty \hskip-5pt dT \, e^{(1-e^T) L^+}  e^{TL}  A_1 * e^{-(e^T-1) L^+_R}~  e^{TL} A_2 * \ldots 
*e^{-(e^T-1) L^+_R}~  e^{TL}A_k  \nonumber \\
&=  & \int_1^\infty \hskip-2pt {ds\over s} \, e^{(1-s) L^+}  s^{L}  A_1 * e^{-(s-1) L^+_R}~  s^{L} A_2 * \ldots 
*e^{-(s-1) L^+_R}~  s^{L}A_k\,.\nonumber 
\end{eqnarray}
Therefore we  have: 
\be
\label{1/L*_on_product}
{1\over L^\star} (A_1 * A_2 * \ldots *A_k) =\int_1^\infty \hskip-2pt {ds\over s}
 \, W_{s-1} * s^{L}  A_1 * W_{s-1} * s^{L} A_2 * \ldots 
* W_{s-1} *  s^{L}A_k * W_{s-1}\,.
\ee
It is interesting to compare with (\ref{1/L_on_prod}).  There are 
two  differences. First,  $1/L^*$ produces two extra factors of $W_{s-1}$, one to the left
and one to the right of the sequence of states.  Second, the range of integration
is different.  While $1/L$ induces contractions $(t\leq 1)$, $1/L^*$ induces expansion
$(s\geq 1)$.  In the language of overlap states
\be
\label{1/L*_on_product_ov}
{1\over L^\star} (A_1 * A_2 * \ldots *A_k) =\int_1^\infty \hskip-2pt {ds\over s}
 \, W_{{1\over 2} (s-1)} *[A_1]_s  * [A_2]_s * \ldots 
*[A_k]_s  * W_{{1\over 2} (s-1)} \,. 
\ee
The off-shell states have been expanded  
 and there are 
wedges flanking the result from the left and from the right.

 The calculation of the $B^\star$ action requires a few formulas. Given the
 $B^\star$ action (\ref{derivation_type}) on star products, one needs the action of
 $B^\star$ and $B^\star - B_L^+$ on wedge states, as given 
 in (\ref{bactiononwedges3}, \ref{bactiononwedges4}).
Given the structure of terms in (\ref{1/L*_on_product}) 
we require the evaluation of 
\be
M' \equiv  (B^\star-B^+_L)
 s^L  A * W_{s-1}  ~+ ~(-1)^A   s^LA *  (B^\star -B^+_L)W_{s-1}\,.
\ee
In the first factor we replace $B^\star-B^+_L = B^+_R - B$ and in the second
we use the $B^\star - B_L^+$ action on wedge states,
\be
M'=  (B^+_R - B)
 s^L  A * W_{s-1}  ~+ ~(-1)^A (s-1)\,  s^LA * B^+_LW_{s-1}\,.
\ee
The $B^+_R$ in the first term can be moved to act on $W_{s-1}$ and the
resulting term combines with the second term:
\be
M'=   -
 s^L  BA * W_{s-1}  ~+ ~(-1)^A s\,  s^LA * B^+_LW_{s-1}\,.
\ee
We can now begin the calculation of 
\be
{B^\star \over L^\star} (A_1 *A_2) = \int_1^\infty \hskip-2pt  {ds\over s}  \,
B^\star(  W_{s-1} *\, s^L A_1 * W_{s-1}* s^L A_2 * W_{s-1})\,.
\ee
Since $B^\star W_{s-1} = s B^+_L W_{s-1}$ we get
\be
\begin{split}
{B^\star \over L^\star} (A_1 *A_2) &= \int_1^\infty \hskip-2pt  ds  \,
\Bigl[ ~B^+_L W_{s-1} *\, s^L A_1 * W_{s-1}*s^L A_2 * W_{s-1}\phantom{Biggl(} \\[0.5ex]
&\qquad~~ +(-1)^{A_1}~~ W_{s-1} *\, s^L A_1 *  B^+_L W_{s-1}* s^L A_2 * W_{s-1}\\[0.5ex]
&\qquad~~ +(-1)^{A_1+A_2} W_{s-1} *\, s^L A_1 *  W_{s-1}* s^L A_2 *  B^+_L W_{s-1}\Bigr]
\\[0.6ex]
&\hskip-40pt- \int_1^\infty \hskip-2pt {ds\over s}   \,
W_{s-1} *\Bigl[ \, s^L BA_1 * W_{s-1}* s^L A_2 
+ (-1)^{A_1}\, s^L A_1 * W_{s-1}* s^L BA_2\Bigr] * W_{s-1}\,.
\end{split}
\ee
\begin{figure}
\centerline{\hbox{\epsfig{figure=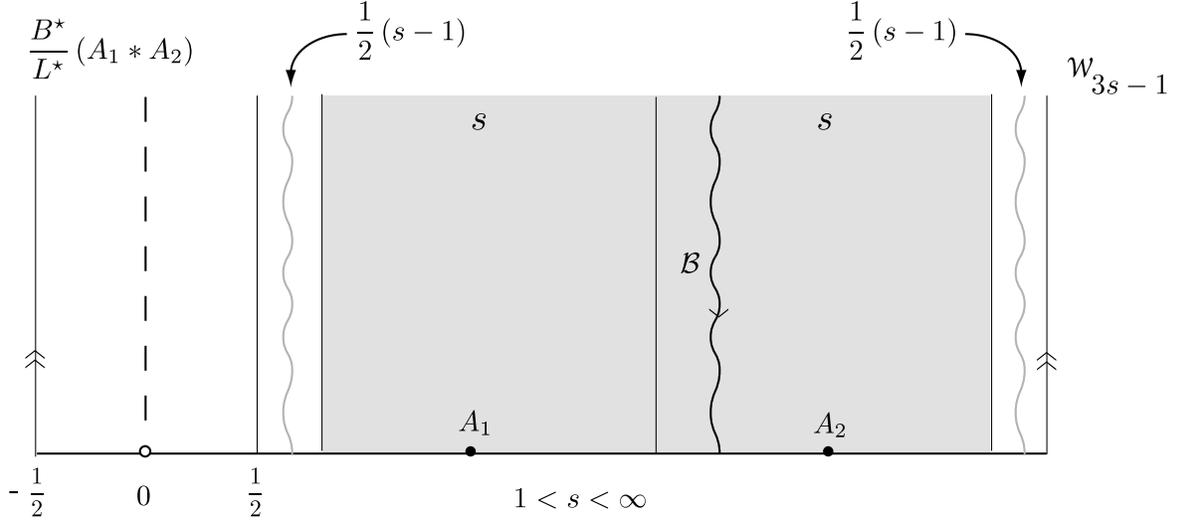, height=6.9cm}}}
\caption{A term in the representation (\ref{ov_not_B*/L*}) 
of ${B^\star\over L^\star} (A_1 *A_2)$.
 Two similar terms terms  in  (\ref{ov_not_B*/L*})  have 
the  $\mathcal{B}$ insertion at the positions shown in gray. The two remaining terms have no $\mathcal{B}$ but rather local insertions of $BA_1$ or $BA_2$. 
Wedges of width ${1\over 2} (s-1)$, with $s\in [1,\infty)$, flank $[A_1]_s *[A_2]_s$.}
\label{rz3fig}
\end{figure}
In the overlap notation, we have 
\be
\label{ov_not_B*/L*}
\boxed{~
\begin{split}
{B^\star \over L^\star} (A_1 *A_2) &= \int_1^\infty \hskip-2pt  ds  \,
\Bigl[ ~B^+_L W_{{1\over 2}(s-1)} *\,[A_1]_s * [A_2]_s  * W_{{1\over 2}(s-1)}\phantom{Biggl(} \\[0.5ex]
&\qquad~~ +(-1)^{A_1}~~ W_{{1\over 2}(s-1)} *\, [A_1]_s *  B^+_L [A_2]_s * 
W_{{1\over 2}(s-1)}\\[0.5ex]
&\qquad~~ +(-1)^{A_1+A_2} W_{{1\over 2}(s-1)} *\, [A_1]_s *  [A_2]_s *  B^+_L W_{{1\over 2}(s-1)}\Bigr]
\\[0.6ex]
&- \int_1^\infty \hskip-2pt {ds\over s}   \,
\Bigl[ W_{{1\over 2}(s-1)} *\,[BA_1]_s * [A_2]_s * W_{{1\over 2}(s-1)}\\[0.5ex]
&\qquad~~+ (-1)^{A_1} W_{{1\over 2}(s-1)} *\, [A_1]_s * [BA_2]_s * W_{{1\over 2}(s-1)}\Bigr]\,.\phantom{\Biggl(}
\end{split}
~}
\ee
The second line on the right-hand side is illustrated in Figure~\ref{rz3fig}.

\sectiono{The off-shell Veneziano amplitude: first two diagrams}

The four point amplitude is obtained by joining two cubic vertices with
a propagator. For arbitrary states $\Psi_i$  ordered as $1234$ along
the boundary, the $\s$-channel contribution is given by 
\be \label{F4}
{\cal F}_\s =   \Bigl\langle \Psi_1 * \Psi_2 ~, {\cal P} ~ (\Psi_3 * \Psi_4) \Bigr\rangle  \,.
\ee
Since ${\cal P}$ is BPZ even, 
${\cal F}_\s$ is symmetric
under the exchange of $(\Psi_1, \Psi_2) \leftrightarrow (\Psi_3, \Psi_4)$. 
One's first instinct is to process $\mathcal{F}_\s$ by
moving the BRST operator in ${\cal P}$ to the right,
\be
\label{schnabl_gauge_propagator_naive2}
\mathcal{P} ~\overset{?}{=} ~\frac{B}{L} -  \frac{B}{L}\frac{B^\star}{L^\star}  \,Q\,\, .
\ee
Then
\be 
\mathcal{F}_\s  ~\overset{?}= \mathcal{F}^{(1)}+\mathcal{F}^{(2)} 
+\mathcal{F}^{(2')} \, ,
\ee
with
\begin{equation}
\label{diagrams_split}
\begin{split}
\mathcal{F}^{(1)}   &\equiv  \Bigl\langle \Psi_1 * \Psi_2 ~,~ \frac{B}{L}  (\Psi_3 * \Psi_4) \Bigr\rangle \, ,  \\
\mathcal{F}^{(2)} & \equiv -\Bigl\langle   \Psi_1 * \Psi_2  ~,~    \frac{B}{L}  \frac{B^\star}{L^\star}  
(Q\Psi_3 * \Psi_4) \Bigr\rangle  \, , \\
\mathcal{F}^{(2')} & \equiv \Bigl\langle   \Psi_1 * \Psi_2  ~,~  \frac{B}{L} \frac{B^\star}{L^\star}  ( \Psi_3 * Q\Psi_4) \Bigr\rangle  \, .
\end{split}
\end{equation}
In this section we evaluate ${\cal F}^{(1)}$ (the ``first diagram'') and ${\cal F}^{(2)} + 
{\cal F}^{(2')}$ (the ``second diagram'') for external off-shell tachyons.
We shall find that 
${\cal F}^{(1)} +{\cal F}^{(2)} + 
{\cal F}^{(2')}$  is not symmetric under the exchange $12 \leftrightarrow 34$ --
in contradiction with the symmetry of the starting point (\ref{F4}).
To the rescue will come a boundary term, discussed at length in the next section.

As usual, we will use the Mandelstam variables:
\be
\s = - (p_1+p_2)^2  \,, ~ \t = - (p_2+ p_3)^2  \,, ~ \u = - (p_1 + p_3)^2 \,, \quad 
\s+ \t + \u = - \sum p_i^2 \,.
\ee

\subsection{First diagram}\label{firstdiagram}

\begin{figure}
\centerline{\hbox{\epsfig{figure=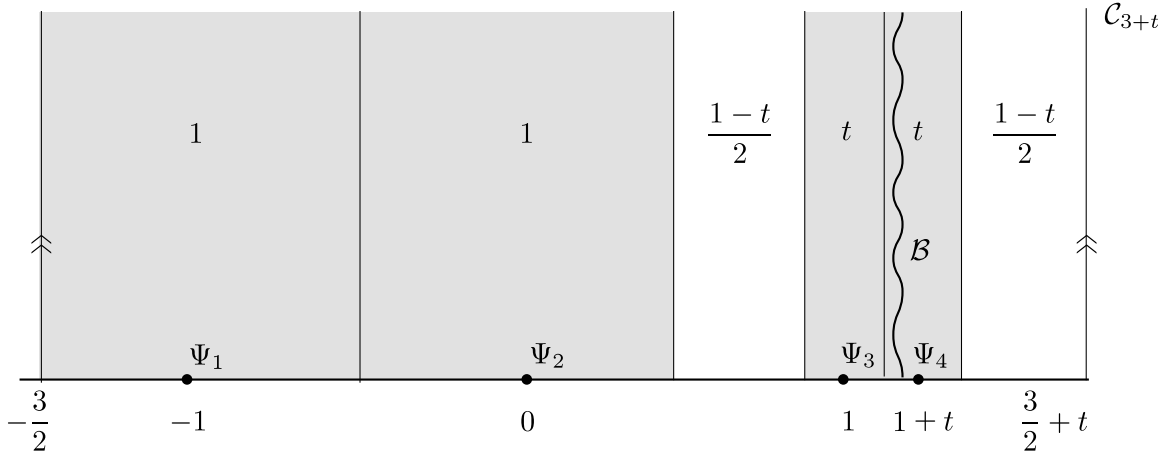, height=6.0cm}}}
\caption{The first string diagram $\mathcal{F}^{(1)}$, with 
$t\in [0,1]$. For $t=0$ punctures 3 and 4 collide. For $t=1$  the punctures
are uniformly spaced on the boundary.}
\label{rz4fig}
\end{figure}

Using (\ref{the_B?Laction}), recalling that $(-B^+_L)$ is represented by the
insertion of $\mathcal{B}$, and noting that the external states are all annihilated
by $B$,  the first diagram $\mathcal{F}^{(1)} $ is given by  
\be
\mathcal{F}^{(1)}   = - \int_0^1 dt\,~\Bigl\langle [\Psi_1]_1 * [\Psi_2]_1 ~,~  W_{{1\over 2} (1-t)} *  [\Psi_3]_t *  \mathcal{B}
 [\Psi_4]_t * W_{{1\over 2} (1-t)} \Bigr\rangle\,.
\ee
The string diagram is the cylinder of total width $3+t$  shown in Figure~\ref{rz4fig}.
Note that states 1 and 2 appear on wedges of unit width, while states 3 and
4 appear on wedges of width~$t$.
We consider general external states of the form
\be
|\Psi_i \rangle = V_i c_1 |0 \rangle \, , 
\ee
where $V_i$ are matter primary operators of dimension $\Delta_i$. We have
\be
L  | \Psi_i \rangle= (\Delta_i -1)| \Psi_i \rangle \, ,\quad Q | \Psi_i \rangle= (\Delta_i -1) V_i c_0 c_1 | 0 \rangle \,.
\ee
The conformal map of $\Psi_i$ by 
\be
z = t \,{2\over \pi} \tan^{-1}(\xi)
\ee
implies that $[\Psi_i]_t$ is represented by the local insertion of 
${\left( 2 t \over \pi \right)^{\Delta_i-1}} \, \Psi_i(z_i)\,.$
Thus we find
\be
\label{fdiag_one}
\mathcal{F}^{(1)}   = - \Bigl({\pi\over 2}\Bigr)^{4- \Delta_i} \int_0^1 {dt\over t^{2-\Delta_3 -\Delta_4}} \,~
\Bigl\langle  c V_1(-1)  \, c V_2(0)   \,c V_3(1)\, \, \mathcal{B}\, c V_4(1+t)  \,    \Bigr\rangle_{{\cal C}_{3+ t}} \, .
\ee
The CFT correlator factorizes into a matter part times a ghost part. It is convenient to use cyclicity and the identification
$z \sim z + 3 +t$ to  rewrite the ghost correlator as
\be
\Bigl\langle c(-1) \, c(0) \, 
  \,c(1)\, \mathcal{B}\, c(1+t)  \,   \Bigr\rangle_{{\cal C}_{3+ t}}  = \Bigl\langle 
\mathcal{B}\, c(-2)\,c(-1) \,c(0)\,  c(1)  \,   \Bigr\rangle_{{\cal C}_{3+ t}} \, .
\ee  
This is immediately evaluated using (\ref{Bccc}). Making also use of the
trigonometric identity $3\sin x - \sin(3x) = 4 \sin^3x$ we find 
\be
\Bigl\langle 
\mathcal{B}\, c(-2)\,c(-1) \,c(0)\,  c(1)  \,   \Bigr\rangle_{{\cal C}_{3+ t}}
=  -\,{4\over \pi^3} \, (3+t)^2\,\sin \Bigl[ {2\pi \over 3+t}\Bigr] \sin^4 \Bigl[ {\pi \over 3+t}
\Bigr]\,.
\ee
To evaluate the matter correlator, we specialize to tachyon vertex operators:  
\be
V_i = e^{\alpha' p_i X} \,, \quad \Delta_i = \alpha' p_i^2 \, .
\ee
Using  (\ref{matter4pt}) we find that
\be
\begin{split}
&\Bigl \langle e^{ip_1\cdot X(-1)} e^{ip_2\cdot X(0)} 
e^{ip_3\cdot X(1)}   e^{ip_4\cdot X(1+t)}   \Bigr\rangle_{{\cal C}_{3+ t}} = 
(2\pi)^D \delta \Bigl(\sum p_i\Bigr)
\Bigl( {3+t\over \pi}\Bigr)^{-\alpha'\sum p_i^2} \\
 &\quad \cdot
  \Bigl( \sin \Bigl[ {\pi\over 3+ t} \Bigr] \Bigr)^{-\alpha' (2\t +\s +p_1^2 + p_2^2 +\sum p_i^2)}
 \Bigl( \sin \Bigl[ {2\pi\over 3+ t} \Bigr] \Bigr)^{\alpha' (2\s+2\t+ \sum p_i^2)}  
  \Bigl( \sin \Bigl[ {3\pi\over 3+ t} \Bigr] \Bigr)^{-\alpha' (\s+ p_4^2+  p_3^2)}\,.
\end{split}
\ee
Assembling our results back into (\ref{fdiag_one}) we have
\be \label{Fs}
\begin{split}
\mathcal{F}^{(1)}   = {4\over \pi}\Bigl({\pi\over 2}\Bigr)^{4- \alpha'\sum p_i^2} 
&(2\pi)^D \delta \Bigl(\sum p_i\Bigr)\int_0^1 {dt\over t^{2-\alpha'(p_3^2 + p_4^2)}}  \, 
\Bigl( {3+t\over \pi}\Bigr)^{2-\alpha'\sum p_i^2}
\, \Bigl( \sin \Bigl[ {3\pi\over 3+ t} \Bigr] \Bigr)^{-\alpha' (\s+ p_4^2+  p_3^2)} \\[0.5ex]
&\hskip-20pt 
\cdot \Bigl( \sin \Bigl[ {\pi\over 3+ t} \Bigr] \Bigr)^{4-\alpha' (2\t +\s +p_1^2 + p_2^2 +\sum p_i^2)}
\cdot \Bigl( \sin \Bigl[ {2\pi\over 3+ t} \Bigr] \Bigr)^{1+\alpha' (2\s+2\t+ \sum p_i^2)}   \,.
 \end{split}
\ee

\begin{figure}
\centerline{\hbox{\epsfig{figure=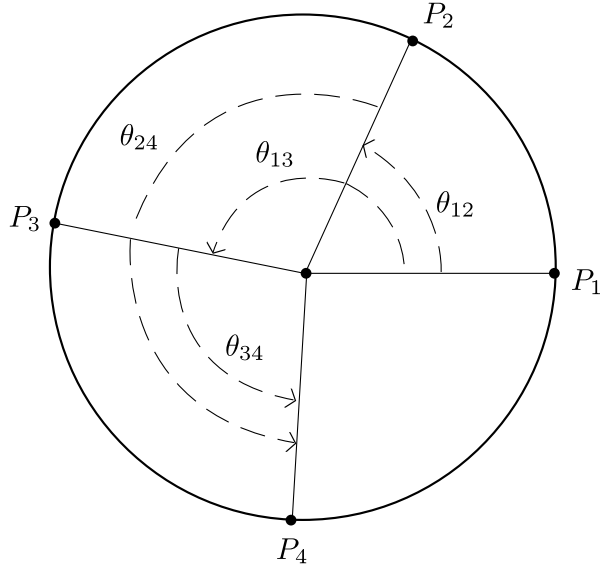, height=7.5cm}}}
\caption{A disk with four punctures $P_i$, $i= 1,2,3,4$.
The angle $\theta_{ij}$, with $i<j$ is defined as the (lowest)
clockwise rotation angle that takes $P_i$ to $P_j$.}
\label{rz9fig}
\end{figure}

It is useful to make a change of variables from $t$ to a more relevant
variable -- the modulus $\lambda$ of the four-punctured disk.
Let us then review how 
to calculate  the modulus $\lambda$ of a disk with four punctures
$P_1, P_2, P_3$, and $P_4$ located with clockwise ordering
on the boundary of the disk.  The modulus $\lambda$, 
with $0\leq \lambda \leq 1$, is defined as the coordinate
of $P_2$ after a map to the upper-half plane in which
$P_1, P_2,$ and $P_3$ go to $0, 1$, and $\infty$, respectively.
Our disk is presented as a circular unit disk
and the relevant information is the angular separations
between the punctures.  We introduce the angle variable
$\theta_{ij}$ with $i< j$ to denote the positive angle of rotation
that is needed to go from  $P_i$ to $P_j$.  The relevant
configuration is shown in Figure~\ref{rz9fig}.  With $w$ denoting
the coordinate on the disk, a map to the upper-half plane is
\be
z  = {1\over i} \, {w-1\over w+1} \,.
\ee
One can readily verify that for points $w= e^{i\theta}$ on the boundary
of the disk
\be
z (w = e^{i\theta}) =  \tan\, {\theta\over 2}\,.
\ee
Placing $P_1$ at $w=1$, the punctures $P_i$ are mapped to $z_i$ where
\be
\label{4ptsmap}
z_1 = 0 \,, \quad z_2 = \tan\, {\theta_{12}\over 2}, \quad
z_3 = \tan\, {\theta_{13}\over 2}\,, \quad z_4 = \tan\, {\theta_{14}\over 2} \,.
\ee
The modulus $\lambda$ is then
\be
\lambda = { (z_1 - z_2) (z_3- z_4)\over (z_1-z_3) (z_2 - z_4)} =
{ z_2 (z_3- z_4)\over z_3 (z_2 - z_4)}\,.
\ee
A short calculation using the values indicated in (\ref{4ptsmap}) gives
the final result
\be
\label{modulus_on_circle}
\lambda ~= ~ {\sin\,\bigl( {\theta_{12}\over 2}\bigr)\, \sin\,\bigl( {\theta_{34}\over 2}\bigr)\over
\sin\,\bigl( {\theta_{13}\over 2}\bigr)\, \sin\,\bigl( {\theta_{24}\over 2}\bigr)} \,.
\ee
To apply this result to cylinder diagrams we note that the angle between
two punctures is simply given by $2\pi$ times the ratio of the separation
between the punctures and the total circumference.  

We apply (\ref{modulus_on_circle}) to the first string diagram (Figure~\ref{rz4fig})
and  find that the modulus $\lambda$ is the given by 
\be
\label{mod_diag1}
\lambda ={ \sin\bigl[ {\pi\over 3+ t} \bigr] \sin\bigl[ {\pi t\over 3+ t} \bigr] \over
\sin^2\bigl[ {2\pi\over 3+ t} \bigr] } \, .
\ee
We readily check that this result  correctly implies that $\lambda (t=0) = 0$,
since then punctures 3 and 4 collide, which is conformally 
equivalent to the collision
of punctures 1 and 2.  We also check that $\lambda (t=1) = 1/2$,
which corresponds to the configuration of equally
spaced punctures on the boundary of the disk.
Using trigonometric identities we find the alternative useful formulas
\be
\label{toflambda}
4\sin^2 \Bigl[ {\pi\over 3+ t} \Bigr]  = {3-4\lambda\over 1-\lambda} \,, 
\qquad 
\cos^2 \Bigl[ {\pi\over 3+ t} \Bigr]  ={1\over 4} \cdot  {1\over 1-\lambda} \,, 
\ee
as well as the Jacobian
\be
{4\over \pi}  {\pi^2\over (3+ t)^2}
\sin \Bigl[ {2\pi\over 3+ t} \Bigr]\,  dt  = {d\lambda\over (1-\lambda)^2} \,,
\ee
which also implies that for small $\lambda$ and small $t$ we have
\be
\label{timittt}
{\lambda\over t} \to  {2\pi\over 3\sqrt{3}} \,.
\ee
Using $\lambda$ as integration variable the full amplitude (\ref{Fs}) becomes 
\be
\begin{split}
\mathcal{F}^{(1)}   = 
&(2\pi)^D \delta \Bigl(\sum p_i\Bigr)\int_0^{1/2} d\lambda  (1-\lambda)^{-2} \, 
\, \Bigl( \sin \Bigl[ {3\pi\over 3+ t} \Bigr] \Bigr)^{-\alpha' (\s+ p_4^2+  p_3^2)} \\[0.5ex]
&\qquad \qquad
\cdot \Bigl( \sin \Bigl[ {\pi\over 3+ t} \Bigr] \Bigr)^{4-\alpha' (2\t +\s +p_1^2 + p_2^2 +\sum p_i^2)}
\cdot \Bigl( \sin \Bigl[ {2\pi\over 3+ t} \Bigr] \Bigr)^{\alpha' (2\s+2\t+ \sum p_i^2)} \\[0.5ex]
&\qquad \qquad\cdot\Bigl({\pi\over 2}\Bigr)^{4- \alpha'\sum p_i^2} 
 \Bigl( {3+t\over \pi}\Bigr)^{4-\alpha'\sum p_i^2}{1\over t^{2-\alpha'(p_3^2 + p_4^2)}} \, .
 \end{split}
\ee
Making use of the identities
\be
\label{identi_diag_one}
\begin{split}
\sin \Bigl[ {\pi\over 3+ t} \Bigr] &=  (1-\lambda)^{-1/2}  \bigl(
{\textstyle {3\over 4}} - \lambda)^{1/2}\,,\\
 \sin \Bigl[ {2\pi\over 3+ t} \Bigr] &=  (1-\lambda)^{-1}  \bigl(
{\textstyle {3\over 4}} - \lambda)^{1/2}\,,\\
 \sin \Bigl[ {3\pi\over 3+ t} \Bigr] &=  \lambda (1-\lambda)^{-3/2}  \bigl(
{\textstyle {3\over 4}} - \lambda)^{1/2}\,,
\end{split}
\ee
a calculation gives the relatively simple result
\be
\label{first_term_ven_amp}
\begin{split}
\mathcal{F}^{(1)}   = ~\,
&(2\pi)^D \delta \Bigl(\sum p_i\Bigr)\int_0^{1/2} d\lambda\, \lambda^{-\alpha' \s -2}  (1-\lambda)^{-\alpha' \t -2} \,  \\[0.5ex]
&\quad\cdot
 \Bigl[{\textstyle {1\over 2}} (3+t) \sin \bigl[ {2\pi\over 3+t} \bigr] \Bigr]^{\chi}\,
\Bigl[ {\lambda \over t\,}\Bigr]^{\chi_{34}}
( 1-\lambda)^{\chi_{12}} \,,
 \end{split} 
\ee
where we have defined variables $\chi_{ij}$  and $\chi$ that vanish on-shell:  
\be \label{chidef}
\boxed{\phantom{\Biggl(}
\chi_{ij} \equiv 2- \alpha'(p_i^2 + p_j^2)  \, ,\quad \chi = 4 - \alpha'\sum_{i=1}^4 p_i^2\,. 
\phantom{\Biggl(} }
\ee
In (\ref{first_term_ven_amp}) one views $t$ as a function of $\lambda$, as 
defined by (\ref{toflambda}).   An alternative expression is
\be
\label{first_term_ven_amp_v2}
\boxed{\phantom{\Biggl(}
\begin{split}
\mathcal{F}^{(1)}   = ~\,
&(2\pi)^D \delta \Bigl(\sum p_i\Bigr)\int_0^{1/2} d\lambda\, \lambda^{-\alpha' \s -2}  (1-\lambda)^{-\alpha' \t -2} \,  \\[0.5ex]
&\quad\cdot
\Bigl[ {\pi\over 2} (1- \lambda)^{1/2}\cdot{3+t\over \pi}
 \sin \bigl[ {\pi\over 3+t} \bigr] \Bigr]^{\chi}\,
\Bigl[ {\lambda \over t\,(1-\lambda)}\Bigr]^{\chi_{34}}\,.\phantom{\Biggl(}
 \end{split}}
\ee
Note that on-shell the second line in the above result gives a factor
of one. The first line then gives us the familiar on-shell Veneziano amplitude.
Of course, one must still add the contribution from the $\t$-channel diagram
to obtain the region of integration $\lambda \in [0,1]$ and then 
the contribution
from other cyclic orderings of the punctures.  Since we have shown that this
first diagram gives the correct on-shell amplitude, it is  clear that the contributions
from the remaining diagrams should vanish on-shell.  

We can use (\ref{first_term_ven_amp_v2}) to obtain the form of the
amplitude near the pole at $\alpha' \s + 1 =0$.  For this we simply 
expand for $\lambda$ near zero and use (\ref{timittt}) to obtain
\be
\label{first_term_pole_amp}
\mathcal{F}^{(1)}|_{pole}   \simeq ~\,
(2\pi)^D \delta \Bigl(\sum p_i\Bigr)\Bigl[ - {1\over \alpha'\s+1} \Bigr]\cdot
 \Bigl[{3\sqrt{3}\over 4}  \Bigr]^{\chi}
\Bigl[{2\pi\over 3\sqrt{3}} \Bigr]^{\chi_{34}}\,,
\ee
or just
\be
\label{first_term_pole_amp_final}
\mathcal{F}^{(1)} |_{pole}  \simeq ~\,
(2\pi)^D \delta \Bigl(\sum p_i\Bigr)\Bigl[ - {1\over \alpha'\s+1} \Bigr]\cdot
 \Bigl[{3\sqrt{3}\over 4}  \Bigr]^{\chi_{12}}
\Bigl[{\pi\over 2} \Bigr]^{\chi_{34}}\,. 
\ee
The external states 1 and 2 appear with the expected  
off-shell factor of the Witten vertex.  States 3 and 4, which collide
 in the string diagram, carry a different off-shell factor.  This
 happens because $B/L$ is not BPZ symmetric. 

We can now ask:  To include the contribution from the $\t$-channel diagram
can we simply extend the region of integration in (\ref{first_term_ven_amp_v2})
from $\lambda \in [0, 1/2]$ to $\lambda \in [0,1]$? Though 
 this would work  on-shell, it does not work off-shell.  In fact, the second line in 
(\ref{first_term_ven_amp_v2}) does not make sense beyond $\lambda = 3/4$,
as can be seen from the top equation in (\ref{identi_diag_one}).  The value
$\lambda = 3/4$ corresponds to the maximum modulus that can be attained in this
string diagram: the modulus for  $t\to \infty$.  

The $\t$-channel answer is obtained by noting that the corresponding diagram
arises from (\ref{F4}) by the replacements  
$\Psi_1 \to \Psi_2$,  
$\Psi_2 \to \Psi_3$,  $\Psi_3\to \Psi_4$ and $\Psi_4 \to \Psi_1$.  This has the
effect of exchanging $\s$ and $\t$ and replacing $\chi_{ij}\to \chi_{i+1, j+1}$  
(understanding that subscripts are defined mod 4).  
We rewrite the $\s$-channel contribution (\ref{first_term_ven_amp_v2}) 
as 
\be
\mathcal{F}^{(1)}_\s   = (2\pi)^D \delta \Bigl(\sum p_i\Bigr)\int_0^{1/2} d\lambda\, \lambda^{-\alpha' \s -2}  (1-\lambda)^{-\alpha' \t -2} \,  
\Bigl[h_1(\lambda) \Bigr]^{\chi}\,
\Bigl[h_2(\lambda) \Bigr]^{\chi_{34}}\,,
\ee
adding the subscript $\s$ to denote that this is the $\s$-channel contribution
and introducing functions $h_1$ and $h_2$ that can be easily read
from (\ref{first_term_ven_amp_v2}).
The contribution from the $\t$-channel would be
\be
\mathcal{F}^{(1)}_\t   = (2\pi)^D \delta \Bigl(\sum p_i\Bigr)
\int_0^{1/2} d\lambda\, \lambda^{-\alpha' \t -2}  (1-\lambda)^{-\alpha' \s -2} \,  
\Bigl[h_1(\lambda) \Bigr]^{\chi}\,
\Bigl[h_2(\lambda) \Bigr]^{\chi_{41}}\,.
\ee
Letting $\lambda \to 1-\lambda$,  we find
\be
\mathcal{F}^{(1)}_\t   = (2\pi)^D \delta \Bigl(\sum p_i\Bigr)
\int_{1/2}^{1} d\lambda\, \lambda^{-\alpha' \s -2}  (1-\lambda)^{-\alpha' \t -2} \,
\Bigl[h_1(1-\lambda) \Bigr]^{\chi}\,
\Bigl[h_2(1-\lambda) \Bigr]^{\chi_{41}}\,.
\ee
Together, $\mathcal{F}^{(1)}_\s$ and $\mathcal{F}^{(1)}_\t$ cover the modular
region $\lambda \in [0,1]$. It is an important consistency condition on the
off-shell amplitude that the integrands of $\mathcal{F}^{(1)}_\s$ and $\mathcal{F}^{(1)}_\t$
match at the boundary point $\lambda = 1/2$.  This matching
occurs for arbitrary value of $h_1(1/2)$  
but requires $h_2(1/2) =1$.  Happily, this
is the case because $\lambda = 1/2$ corresponds to $t=1$.

\subsection{Second diagram}

  The formulas developed so far
are in principle sufficient to evaluate ${\cal F}^{(2)}$ and ${\cal F}^{(2)'}$ in (\ref{Fs}).  Using
BPZ conjugation, we could write
\be
\mathcal{F}^{(2)}  = -\Bigl\langle  \frac{B^\star}{L^\star} ( \Psi_1 * \Psi_2 ) ~,~  \frac{B^\star}{L^\star}  
(Q\Psi_3 * \Psi_4) \Bigr\rangle  \, ,
\ee
and proceed by computing the action of  $B^\star/L^\star$ on the two star products. 
The resulting expression is the sum of ten inequivalent terms and would be
rather lengthy to evaluate.  Instead, we are going to use a simpler
method  inspired by a similar treatment in \cite{japanese}.

\begin{figure}
\centerline{\hbox{\epsfig{figure=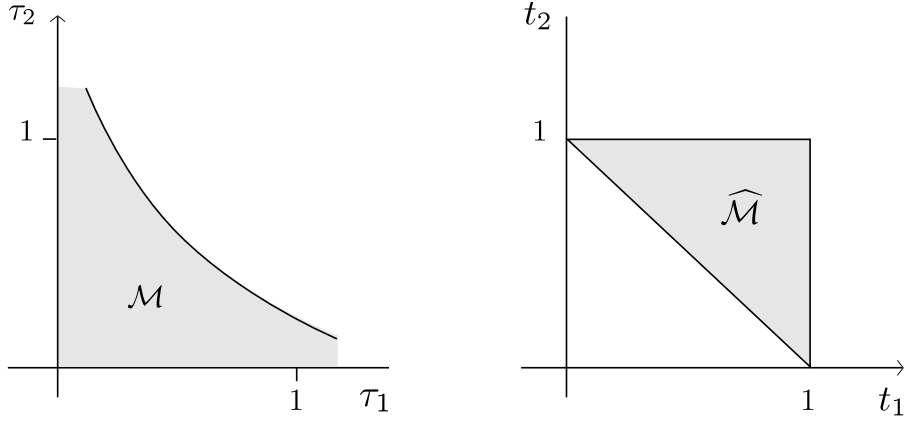, height=5.5cm}}}
\caption{Left: region of integration on the right-hand side of the rearrangement
formula (\ref{trick}).  Right:  The same region of integration in triangular variables.}
\label{rz7fig}
\end{figure}

Since we have seen that
the action of $B/L$ on star products is somewhat simpler than the action
of $B^\star/L^\star$,  it is advantageous to reorder 
\be \label{blbsls}
\frac{B}{L} \frac{B^\star}{L^\star} = \int_0^\infty  dT_1 dT_2 \; B  \, e^{-T_1 L} \, e^{-T_2 L^\star}  B^\star 
\ee
in such a way that the $L$ and $B$ operators are moved to the right of the $L^\star$ and $B^\star$ operators.
In appendix C we prove the elegant rearrangement formula
\be \label{trick}
\boxed{\phantom{\Biggl(}
\frac{B}{L}  \frac{B^\star}{L^\star} =- \int_{{\cal M}} d \tau_1 d \tau_2 \;   B^\star  e^{-\tau_2 L^\star} \; B   \, e^{-\tau_1 L} \, ,}
\ee
where the region of integration ${\cal M}$ is defined by
\be
\boxed{
\phantom{\Biggl(}{\cal M}  \equiv \{ \tau_1 , \tau_2  \; |  \tau_1 \geq 0\,, \tau_2 \geq 0\,, e^{-\tau_1} + e^{-\tau_2} \geq 1 \} \,.}
\ee
The region $\mathcal{M}$ is shown in the left part of Figure~\ref{rz7fig}.  Note that
if the integration were 
 over the whole range of positive $\tau_1$ and $\tau_2$, the
right-hand side of (\ref{trick}) would be equal to $-\frac{B^\star}{L^\star}\frac{B}{L} $.

With the use of the rearrangement formula the contributions (\ref{diagrams_split}) to 
the second diagram can be written as
\begin{eqnarray}
\mathcal{F}^{(2)} &   =  &     \int_{{\cal M} } d \tau_1 d \tau_2 \; 
\Bigl\langle  B e^{- \tau_2 L}  ( \Psi_1 * \Psi_2 ) ~,~  B e^{-\tau_1 L}
(Q\Psi_3 * \Psi_4) \Bigr\rangle  \, . \\
\mathcal{F}^{(2')} &   =  &   -  \int_{{\cal M} } d \tau_1 d \tau_2 \; 
\Bigl\langle  B e^{- \tau_2 L}  ( \Psi_1 * \Psi_2 ) ~,~  B e^{-\tau_1 L}
(\Psi_3 * Q \Psi_4) \Bigr\rangle  \, .
\end{eqnarray}
Evaluating the action of $B e^{-\tau L}$
on the star products by the techniques of \S2,  and using $t_1 = e^{-\tau_1}$,  $t_2 = e^{-\tau_2}$, we find
\begin{eqnarray}
\label{inoverlap2nd}
 && \mathcal{F}^{(2)} =  \int_{\widehat {\cal M}} dt_1 dt_2 \;   \Bigl\langle \, W_{\frac{1}{2}(1-t_2)} * [ \Psi_1]_{t_2} * B^+_L [\Psi_2]_{t_2} *  W_{\frac{1}{2}(1-t_2)}\, {\bf ,} \\   &&  -  W_{\frac{1}{2}(1-t_1)} * [Q \Psi_3]_{t_1} * B^+_L [\Psi_4]_{t_1} * W_{\frac{1}{2}(1-t_1)}+\frac{1}{t_1}  
W_{\frac{1}{2}(1-t_1)} * [ B Q \Psi_3]_{t_1} * [\Psi_4]_{t_1} * W_{\frac{1}{2}(1-t_1)} \, 
\Bigr\rangle\, ,\nonumber
\end{eqnarray}
where the region of integration is the triangular region
\be
\widehat {\cal M} \equiv \{ t_1 , t_2  \; | ~ 0 \leq t_1 \leq 1, ~~0 \leq t_2 \leq 1\,,
~~ t_1 + t_2 \geq 1 \} \,,  
\ee
shown on the right part of Figure~\ref{rz7fig}.
The string diagram for the first contribution in (\ref{inoverlap2nd}) is shown
as a cylinder of  width $2+t_1+t_2$ in Figure~\ref{rz8fig}. 
\begin{figure}
\centerline{\hbox{\epsfig{figure=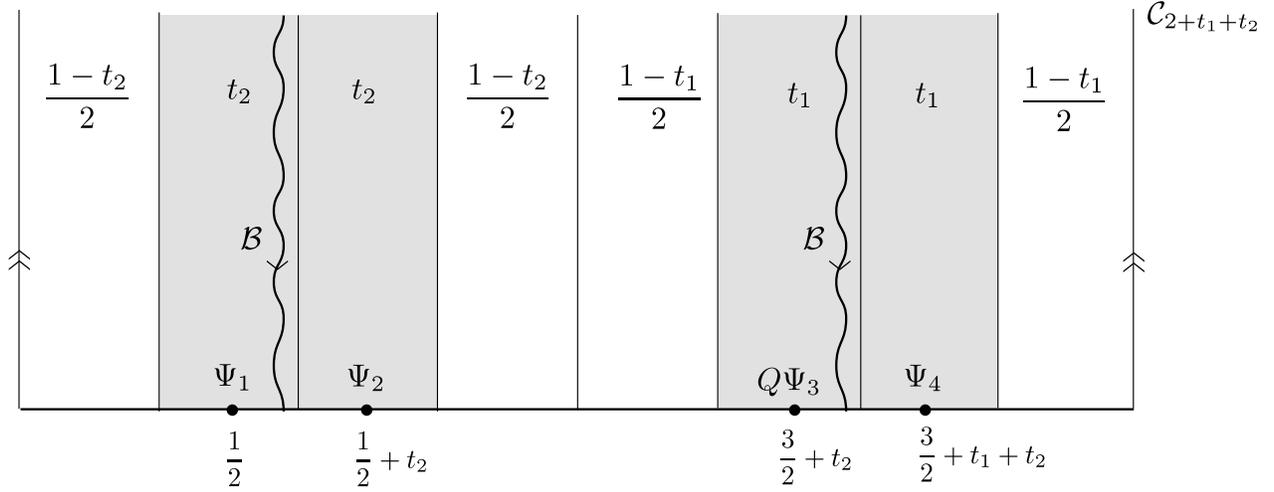, height=6.5cm}}}
\caption{One of the terms in the second string diagram.}
\label{rz5fig}
\end{figure}

\begin{figure}
\centerline{\hbox{\epsfig{figure=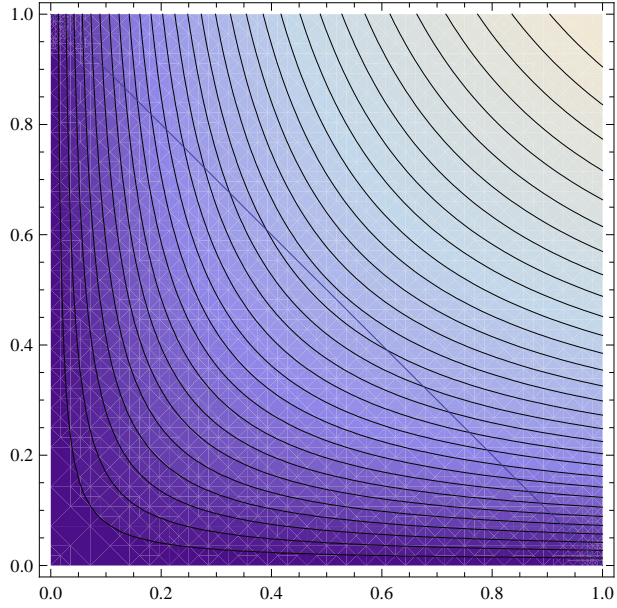, height=8.0cm}}}
\caption{Curves of constant modulus $\lambda$ on the $t_1$, $t_2$
unit square.  The triangle above the diagonal is the region
of integration $\widehat{\mathcal{M}}$  for the second string diagram
(see Figure~\ref{rz7fig}). }  
\label{mstfig}
\end{figure}

Taking  states  of the from $\Psi_i = c V_i$, we find
\be
\mathcal{F}^{(2)} = -   \int_{\widehat {\cal M}} dt_1 dt_2 \;   \left( 2 t_2 \over \pi \right)^{\Delta_1 + \Delta_2-2} \left( 2 t_1 \over \pi \right)^{\Delta_3 + \Delta_4-2}\, (\Delta_3-1)\,
 A^m \Bigl(\frac{A^g_1}{t_1} + A^g_2\Bigr) \, ,
\ee
where we have defined the following matter and ghost correlators,
\begin{eqnarray}
 A^m  & \equiv &  \langle V_1 (r_1) V_2(r_2) V_3 (r_3) V_4 (r_4)
  \rangle_{{\cal C}_\ell}\,, \\
A^g_1  & \equiv &  \langle c (r_1) {\cal B} c(r_2) c (r_3) c (r_4) 
\rangle_{{\cal C}_\ell} \,,\\
A^g_2  & \equiv  &   \langle c (r_1) {\cal B} c(r_2) \partial c c (r_3) {\cal B} c (r_4) \rangle_{{\cal C}_\ell}\, .
\end{eqnarray}
Here the insertion points and the circumference of the cylinder are given by
\be
 r_1 = \frac{1}{2}\, ,\quad  r_2= \frac{1}{2} + t_2\, , \quad r_3 = \frac{3}{2} + t_2 \, ,\quad r_4 = \frac{3}{2} + t_2 + t_1 \, ,\quad l = t_1 +t_2 +2 \
\, .\ee
As before, the evaluation of the ghost correlators is simplified by using cyclicity and the periodic identification to write
\begin{eqnarray}
A^g_1  & = &  \langle {\cal B} c(r_2) c (r_3) c (r_4) c (r_1+ l) \rangle_{{\cal C}_\ell} \, , \\
A^g_2  & =  &   \langle  {\cal B} c(r_2) \partial c c (r_3) {\cal B} c (r_4) c (r_1+ l) \rangle_{{\cal C}_\ell} .
\end{eqnarray}
It is convenient to introduce the shorthand 
\be
\gamma \equiv  \frac{\pi}{\ell} = \frac{\pi}{2 + t_1 + t_2}\,.
\ee
Using the formulas of appendix B, a calculation gives
\be
\begin{split}
A_1^g &= -\frac{1} { \pi \gamma^2}   \sin(\gamma) \sin(\gamma (t_1+1))\, \left[(t_1+2) \sin(\gamma t_1) - t_1 \sin (\gamma(t_1 +2)\right] \,,\\[1.0ex]
A_2^g ~ &= ~~~{1\over \pi \gamma^2} \,\sin(\gamma) \sin( \gamma (t_1+1))
\bigl( \sin (\gamma t_1) - \sin (\gamma t_2) \bigr) \\[1.0ex]
&~~~~+  {1\over \pi \gamma} \,
\bigl[ \sin^2 (\gamma) - \sin^2 (\gamma t_1) + \sin^2 (\gamma (t_1+1)) \bigr]\,.
\end{split}
\ee
Combining terms we find
\be
\frac{A^g_1}{t_1} + A^g_2 =   {1\over \pi \gamma} \,
\bigl( \cos (\gamma t_1) +\cos (\gamma t_2)\bigr) \Bigl( \cos (\gamma t_1) - 
{\sin (\gamma t_1) \over \gamma t_1} \Bigr) \,.
\ee
An entirely analogous calculation gives
\be
\mathcal{F}^{(2')} = -   \int_{\widehat {\cal M}} dt_1 dt_2 \;   \left( 2 t_2 \over \pi \right)^{\Delta_1 + \Delta_2-2} \left( 2 t_1 \over \pi \right)^{\Delta_3 + \Delta_4-2}\, (\Delta_4 -1) \,
 A^m \Bigl(\frac{A^g_1}{t_1} + A^g_3\Bigr) \, ,
\ee
where
$A^g_3  \equiv   
 \langle c (r_1) {\cal B} c(r_2) c (r_3) {\cal B} \partial c c (r_4) \rangle_{{\cal C}_\ell}$.
Explicit computation reveals that  $A^g_3 = A^g_2$, so that
\be
\mathcal{F}^{(2)}+\mathcal{F}^{(2')} = -   \int_{\widehat {\cal M}} dt_1 dt_2    \left( 2 t_2 \over \pi \right)^{\Delta_1 + \Delta_2-2}\hskip-5pt \left( 2 t_1 \over \pi \right)^{\Delta_3 + \Delta_4-2}\hskip-5pt (\Delta_3 + \Delta_4 -2) \,
 A^m \Bigl(\frac{A^g_1}{t_1} + A^g_2\Bigr) \, .
\ee
We now specialize to  the case of tachyons $V_i = e^{\alpha' p_i X}$. Evaluating the matter correlator through (\ref{matter4pt})
and collecting all the terms 
we finally find
\be
\label{second_part_amplitude}
\boxed{\phantom{\Biggl(}
\begin{split}
 \mathcal{F}^{(2)} + \mathcal{F}^{(2')}  &= ~ -\, (2\pi)^{D} \delta \Bigl(\sum p\Bigr)\,  \frac{1}{\pi}\left( \frac{2}{\pi}\right)^{\alpha' \sum p^2 - 4}(\alpha' (p_3^2 + p_4^2) -2)\\[0.5ex]
& \qquad \cdot \int_{\widehat {\cal M}} dt_1 dt_2 \; \gamma^3 \left(\gamma t_2 \right)^{\alpha' (p_1^2 +  p_2^2) -2}\left(\gamma t_1  \right)^{\alpha' (p_3^2 +  p_4^2) -2} 
\\&\qquad\qquad~~ \cdot 
(\cos (\gamma t_2 )+ \cos( \gamma t_1) )\Bigl( \cos (\gamma t_1) - {\sin(\gamma t_1)\over \gamma t_1} \Bigr)  \\[0.5ex]
 &\qquad\qquad~~ \cdot 
 \bigl|\sin (\gamma t_1)\bigr|^{-\alpha' (\s + p_3^2 +p_4^2)}
 \cdot
 \bigl|\sin(\gamma t_2 )\bigr|^{  -\alpha' (\s +  p_1^2 +  p_2^2)  } \\[1.2ex]
&\qquad\qquad~~ 
\cdot \bigl| \sin \gamma \bigr|^{- \alpha' (2\t + \sum p^2)}\cdot \bigl|\sin ( \gamma(t_1 +1))\bigr|^{\alpha' (2\s + 2 \t + \sum p^2) }  \,.\\&
\end{split} \phantom{Biggl(} }
\ee
This is in agreement (up to an overall minus sign) with the result quoted in 
\cite{japanese}, as can be checked using the change of variables
\be
T^{there}_1 = \frac{2 t_2}{t_1 + t_2 -1} \, , 
\quad T^{there}_2 = \frac{2 t_1}{t_1 + t_2 -1} \, .
\ee
To understand the result we express it in terms of the modular 
parameter $\lambda$ for the string diagram.  Using
(\ref{modulus_on_circle}) we find  
\be
\label{mod_diag2}
\lambda 
= ~ \frac{\sin(\gamma t_1) \sin(\gamma t_2)}{ \sin^2(\gamma (t_1 +1))  } \,.
\ee
A small computation then gives
\be
\label{mod_diag22}
1-\lambda = ~ { \sin^2(\gamma )\over \sin^2(\gamma (t_1+1))} \,.
\ee
It is interesting now to appreciate how $\lambda$ varies as a function
of $t_1$ and $t_2$. Figure~\ref{mstfig} shows
 the curves of constant
$\lambda$ on the unit square $0\leq t_1, t_2 \leq 1$.  Shown is also the diagonal that defines the
upper right triangle as the relevant region of integration
$\widehat{\mathcal{M}}$.  The curves of $\lambda =0$ are the 
lines $t_2=0$ and $t_1=0$.  The point $t_1 = t_2 = 1/2$ corresponds
to $\lambda = 1/4$, so in fact, all curves of constant $\lambda < 1/4$ intersect
the diagonal twice.  The curves with $\lambda >1/4$ are above the diagonal,
and the point $t_1 = t_2 =1$ corresponds to $\lambda = 1/2$.  

Using the modulus $\lambda$ we can  rewrite the amplitude as   
\be
\label{second_part_amplitude99}
\begin{split}
 \mathcal{F}^{(2)} + \mathcal{F}^{(2')}  
 &= ~ \, (2\pi)^{D} \delta \Bigl(\sum p\Bigr)\,  \frac{1}{\pi} \, 
 \chi_{34 }\\[1.0ex]
& \qquad \cdot \int_{\widehat {\cal M}} dt_1 dt_2 \;  \lambda^{-\alpha' \s -2}  \, (1-\lambda)^{-\alpha' \t -2} 
\\[1.5ex]
&\qquad~\cdot \gamma^3 \,
{(\cos \gamma t_2 + \cos \gamma t_1 )\over \sin^4 \gamma (t_1+1)}\,
\Bigl( \cos \gamma t_1 - {\sin \gamma t_1\over \gamma t_1} \Bigr)  \\[1.0ex]
 &\qquad~ \cdot \Bigl|{\sin \gamma t_2\over \gamma t_2} \Bigr|^{\chi_{12}}
\cdot \Bigl|{\sin \gamma t_1\over \gamma t_1} \Bigr|^{\chi_{34}} 
\cdot \Bigl( {\pi\over 2 } (1-\lambda)^{1/2}\Bigr)^{\chi }
.
\end{split} \ee
To simplify the answer further we introduce the variable 
$u$:
\be \label{udef}
u = {\sin \gamma t_1\over \gamma t_1} \,.
\ee
The motivation for introducing $u$ is that its derivative  with respect
to $\gamma t_1$ produces 
the factor  $\cos \gamma t_1 - {\sin \gamma t_1\over \gamma t_1}$.  Remarkably the full Jacobian gives 
\be
d\lambda\, {du\over u} = {1\over \pi}\, dt_1 dt_2  \,{(\cos \gamma t_2 + \cos \gamma t_1 )\over \sin^4 \gamma (t_1+1)}\,
\Bigl( {\sin \gamma t_1\over \gamma t_1}- \cos \gamma t_1  \Bigr) \,.
\ee
The amplitude then collapses to a relatively simple form:  
\be
\label{second_part_amplitude999}
\boxed{\phantom{\Biggl(}
\begin{split}
 \mathcal{F}^{(2)} + \mathcal{F}^{(2')}  
 &= ~ -\, (2\pi)^{D} \delta \Bigl(\sum p\Bigr)\, 
 \int d\lambda du \;  \lambda^{-\alpha' s -2}  \, (1-\lambda)^{-\alpha' t -2} 
\\[1.5ex]
 &  \qquad~\qquad \cdot \Bigl( {\pi\over 2} (1-\lambda)^{1/2}\Bigr)^{\chi}\cdot \Bigl|{\sin \gamma t_2\over \gamma t_2} \Bigr|^{\chi_{12}}
\cdot\Bigl( {d\over du}  u^{\chi_{34}} \Bigr) 
.\\&
\end{split} \phantom{Biggl(} }
\ee
The region of integration here includes $0\leq \lambda \leq 1/2$ and 
values of $u$ that depend on $\lambda$ -- the details of which have
been anticipated in our discussion of Figure~\ref{mstfig}.  Indeed, $u$ is
a parameter for the curves of constant modulus, and the integration
is restricted over the upper right triangle.  For $0\leq \lambda \leq 1/4$
each curve of constant modulus has two pieces inside the integration domain
and one piece outside of it.  For $1/4\leq \lambda \leq 1/2$
the curves of constant modulus are entirely contained 
inside the integration domain.

The diagram studied in this subsection (the second diagram) 
does not contribute to the
pole at $\alpha' \s +1$, it is a regular function at this kinematic point.
To see this, 
it is easiest to start with the representation (\ref{second_part_amplitude}).  Any singularity must arise because the 
integrand goes to infinity somewhere over the compact  domain of integration. A little thought shows that this can only happen at the corners 
where either $t_1$ or $t_2$ are equal to zero.  In fact, when $t_1$ is equal
to zero the integrand is regular for $\alpha' \s +1=0$.   As $t_2\to 0$ there 
is a candidate singularity.  To explore it, we make the change of variables $(t_1, t_2) \to (\rho,  t_2)$, where
\be
\rho \equiv \frac{t_1-1}{t_2} \,.
\ee
We then have
\be
\int_{\widehat {\cal M} }  dt_1 dt_2 = \int_0^1 d  \rho \int_0^1dt_2 \,  t_2 \, .
\ee
The extra factor of $t_2$, arising from the Jacobian,  makes
the integrand a finite function as $t_2\to 0$ and this 
 establishes
 our claim. While there is no pole for $\alpha'\s = -1$, the amplitude 
 under consideration has a pole for $ \alpha'\s =0$, 
 corresponding to the exchange of on-shell massless states.  There are also poles at positive
 integer values of $\alpha' \s$.

 The failure of (\ref{second_part_amplitude999}) to contribute to the
 pole at $\alpha' \s + 1=0$ means that near that pole the full amplitude
 obtained from the first and second diagrams still behaves as in (\ref{first_term_pole_amp_final}).  We conclude that the full amplitude computed so far
 does not have the expected symmetry under the exchange 
 $(p_1, p_2) \leftrightarrow (p_3 , p_4)$.

\sectiono{Boundary term and exchange symmetry}

The calculation of the previous section has failed to give a result
consistent with the expected symmetry $(p_1, p_2) \leftrightarrow (p_3 , p_4)$.
The error can be traced to the naive manipulation 
(\ref{schnabl_gauge_propagator_naive2}), where a boundary term was
inadvertently dropped. Let us repeat this manipulation more carefully. We introduce a regulated
version of the propagator: 
\be
{\cal P}_{\Lambda \Lambda^\star} \equiv   \frac{B}{L_\Lambda} \, Q  \,\frac{B^\star}{L^\star_{\Lambda^\star}}  \, ,
\ee
where
\be
  \frac{1}{L_\Lambda}  \equiv    \int_0^\Lambda dT_1   \; e^{-T_1 L}  \, , 
  \quad  \frac{1}{L^\star_{\Lambda^\star}}   \equiv    \int_0^{\Lambda^\star} dT_2   \; e^{-T_2 L^\star}  \, .
\ee
Ultimately we are interested in taking the limit $\Lambda\,, \Lambda^\star \to \infty$. Under  BPZ conjugation
 the cutoffs $\Lambda$ and $\Lambda^\star$ get interchanged,  
 \be \label{BPZreg}
 {\cal P}^\star_{\Lambda \Lambda^\star} = {\cal P}_{\Lambda^\star \Lambda} \,.
 \ee
Using $\{ Q, B^\star\} = L^\star$, we have
\be \label{moveQ}
{\cal P}_{\Lambda \Lambda^\star}    =   -     \int_0^\Lambda  dT_1      \int_0^{\Lambda^\star } \hskip-5pt  dT_2 \,B e^{-T_1 L} \, B^\star e^{-T_2 L^\star}  Q
+     \int_0^\Lambda dT_1  \int_0^{\Lambda^\star} \hskip-6pt  dT_2 \,  B e^{-T_1 L}   \left(  -\frac{d}{dT_2}  e^{-T_2 L^\star}  \right)\,,
\ee
which yields
\be
 {\cal P}_{\Lambda \Lambda^\star}=-   \,  \frac{B}{L_\Lambda }   \frac{B^\star}{L^\star_{\Lambda^\star}} \,     Q    +          \frac{B}{L_\Lambda}   - 
   \frac{B}{L_\Lambda}   \,e^{-\Lambda^\star L^\star}  \, .
\ee
The last term is the boundary term that was previously dropped. It gives
an additional contribution to the four point function:  
\be \label{regboundary}
{\cal F}^B_{\Lambda \Lambda^\star} \equiv  - \Bigl \langle   \Psi_1 * \Psi_2,  \, \int_0^\Lambda dT_1    \,  B e^{-T_1 L}  \,e^{-\Lambda^\star L^\star}  
(\Psi_3 * \Psi_4)  \Bigr \rangle \,.
\ee   
It is useful to understand intuitively why the boundary term $\frac{B}{L_\Lambda}   \,e^{-\Lambda^\star L^\star} $ can give a contribution.  Naively one would argue that in the
limit $\Lambda^\star \to \infty$, the factor  $e^{-\Lambda^\star L^\star}$  gives rise to degenerate Riemann surfaces and no regular contribution to amplitudes.  It is the interplay with $\frac{B}{L_\Lambda}$ that invalidates this argument.  While $e^{-\Lambda^\star L^\star}$ expands the surface
by a scale factor of order $e^{\Lambda^\star}$, the factor $\frac{B}{L_\Lambda}$ sums over all surfaces obtained by contraction
with scales that go from one down to about $e^{-\Lambda}$. 
 If one takes the $\Lambda \to \infty$ limit first, we have $\frac{B}{L}   \,e^{-\Lambda^\star L^\star} $, which induces a very large expansion followed by a set of contractions that go down to zero size.  This is a large set of {\em non-degenerate} surfaces.  Had we taken the limit $\Lambda^\star
\to 0$ first we would have indeed found only degenerate surfaces.

The terms previously computed,   ${\cal F}^{(1)}$ and ${\cal F}^{(2)} + {\cal F}^{(2')}$,
  do not depend on the precise way one takes  $\Lambda$ and $\Lambda^\star $ to infinity, in particular
the order of the $\Lambda$ and $\Lambda^\star$ limits can be safely interchanged. This is obvious for the first diagram,
and is also true for  the second diagram, because the integration region $\widehat {\cal M}$ has
an unambiguous limiting shape for large $\Lambda$ and $\Lambda^\star$.
 We define ${\cal F}^B$ as a suitable limit of ${\cal F}^B_{\Lambda\Lambda^\star}$.
As anticipated above, we shall see that ${\cal F}^B$ 
depends on the prescription used to take this limit. We choose
the prescription by requiring that

\vspace{0.2cm}

(i) ${\cal F}^B$ vanishes for external on-shell states;

\vspace{0.2cm}

 (ii) the total amplitude  ${\cal F}_\s = {\cal F}^{(1)} + {\cal F}^{(2)} + {\cal F}^{(2')} + {\cal F}^B$  
is symmetric under 
$(p_1, p_2) \leftrightarrow (p_3, p_4)$.

\vspace{0.2cm}

Property (i) is necessary since
for on-shell tachyons  the naive calculation  already gave the right result.
We claim that the correct prescription is
\be \label{symm_prescr}
\boxed{\phantom{\Biggl(}
{\cal P} \equiv  \frac{1}{2} \Bigl( \lim_{\Lambda \to \infty}  \lim_{\Lambda^\star \to \infty}  + 
 \lim_{\Lambda^\star \to \infty}  \lim_{\Lambda \to \infty} \Bigr) {\cal P}_{\Lambda \Lambda^\star} \,, \quad  {\cal P}_{\Lambda \Lambda^\star} =   \frac{B}{L_\Lambda} \, Q  \,\frac{B^\star}{L^\star_{\Lambda^\star}}  \, ,
 \phantom{\Biggl(} }
\ee
Using (\ref{BPZreg}) we see that this leads to a BPZ even propagator, so
barring calculational errors this prescription must give a four-point amplitude that obeys property (ii). We confirm
this fact in the rest of the section: the exchange symmetry
is restored by the boundary term (\ref{regboundary})
with the limits taken according to (\ref{symm_prescr}). Less obviously, property (i) also holds.
  It is non-trivial that {a} prescription with the right properties exists.  We also re-examine
  work of \S\ref{derprop} and confirm that the propagator $\mathcal{P}$ defined 
  above provides an inverse to the kinetic operator in some appropriate sense.

\subsection{Boundary term}

\begin{figure}
\centerline{\hbox{\epsfig{figure=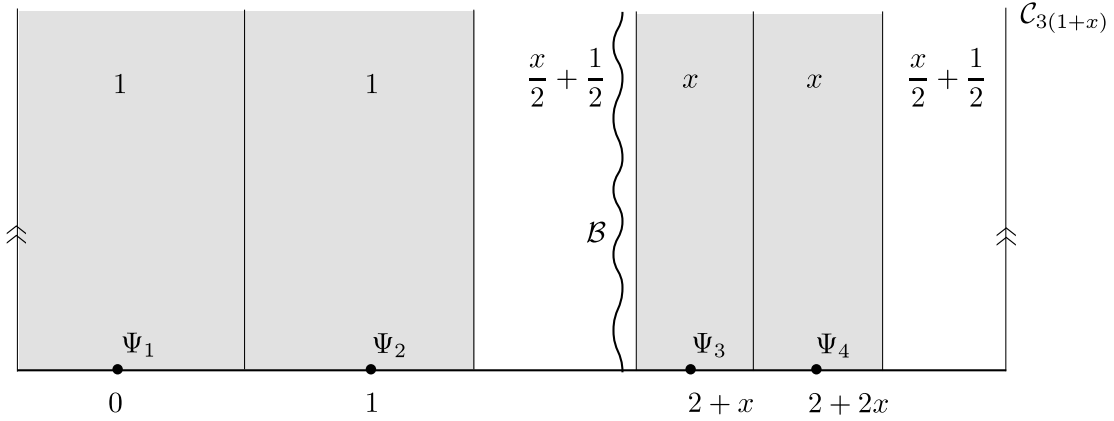, height=5.5cm}}}
\caption{The string diagram for the first term in the 
boundary amplitude $\cal{F}^B$ in (\ref{o3rugue}).}   
\label{rz6fig}
\end{figure}

We now turn to an explicit evaluation of the boundary term.
By the techniques of \S 2, we find
\be \label{FB}
{\cal F}^B_{\Lambda \Lambda^\star} =  - \int_{ \frac{1}{s} }^1 \frac{dt}{t}  \;  \Bigl \langle \Psi_1 * \Psi_2,  B \left( 
W_{\frac{t s^\star}{2} - t + \frac{1}{2} } * [\Psi_3 ]_{ts^\star} *[\Psi_4]_{ts^\star} * W_{\frac{ts^\star}{2} - t + \frac{1}{2} } \right ) \,  \Bigr \rangle \, ,
\ee
where we have set
\be
s \equiv e^{\Lambda} \, , \quad s^\star \equiv e^{\Lambda^\star} \, .
\ee
We evaluate the action of $B$, assuming that $\Psi_3$ and $\Psi_4$ are
 in Schnabl gauge:
\begin{eqnarray}
 && B \left( 
W_{\frac{t s^\star}{2} - t + \frac{1}{2} } 
* [\Psi_3 ]_{ts^\star} *[\Psi_4]_{ts^\star} * W_{\frac{t s^\star}{2} - t + \frac{1}{2} } \right )    =  \\[0.5ex] 
  &&  t (1-s^\star) W_{\frac{t s^\star}{2} - t + \frac{1}{2} } * B^+_L [\Psi_3 ]_{ts^\star} *[\Psi_4]_{ts^\star} * W_{\frac{ts^\star}{2} - t + \frac{1}{2} } \nonumber \\[0.5ex] 
 && -   t s^\star  (-1)^{\Psi_3}   W_{\frac{t s^\star}{2} - t + \frac{1}{2} } *  [\Psi_3 ]_{ts^\star} *B^+_L [\Psi_4]_{ts^\star} * W_{\frac{ts^\star}{2} - t + \frac{1}{2} } \nonumber  \\[0.5ex]
&& + t (1-s^\star)   (-1)^{\Psi_3+ \Psi_4}   W_{\frac{t s^\star}{2} - t + \frac{1}{2} } *  [\Psi_3 ]_{ts^\star} * [\Psi_4]_{ts^\star} *  B^+_L W_{\frac{ts^\star}{2} - t + \frac{1}{2} }\,.\nonumber
\end{eqnarray}
Back in (\ref{FB}) and changing the integration variable to $x = s^\star t$,  
\begin{eqnarray} \label{FBregfinal}
 {\cal F}^B_{\Lambda \Lambda^\star} =   \int_{\frac{s^\star}{s}}^{s^\star} dx 
 &  \Bigl\{    &  (1-\frac{1}{s^\star})\;   \langle \Psi_1 * \Psi_2,  W_{\frac{x}{2} - \frac{x}{s^\star} + \frac{1}{2} } * B^+_L [\Psi_3 ]_{x} *[\Psi_4]_{x} * W_{\frac{x}{2} - \frac{x}{s^\star} + \frac{1}{2} }  \rangle \\
 && +\,  (-1)^{\Psi_3} \;  \langle \Psi_1 * \Psi_2,  W_{\frac{x}{2} - \frac{x}{s^\star} + \frac{1}{2} } * [\Psi_3 ]_{x} * B^+_L [\Psi_4]_{x} * W_{\frac{x}{2} - \frac{x}{s^\star} + \frac{1}{2} }  \rangle \nonumber \\
 &&  + \,(-1)^{\Psi_3+ \Psi_4} (1-\frac{1}{s^\star})  \; \langle \Psi_1 * \Psi_2,  W_{\frac{x}{2} - \frac{x}{s^\star} + \frac{1}{2} } * [\Psi_3 ]_{x} * [\Psi_4]_{x} * B^+_L W_{\frac{x}{2} - \frac{x}{s^\star} + \frac{1}{2} }  \rangle \; \Big\} \, . \nonumber
\end{eqnarray}
We now remove the regulators using the  symmetrized prescription (\ref{symm_prescr}). The first term in (\ref{symm_prescr}) is
an instruction to send $s^\star \to \infty$ first: then both the lower and upper limits of integration in (\ref{FBregfinal}) go to infinity,
and only the singular surface with $x=\infty$ is kept. We believe that this singular surface gives no contribution to the integral --
 the first term in (\ref{symm_prescr})  does not contribute to ${\cal F}^B$. On the other hand, the second term in (\ref{symm_prescr})
 gives a regular contribution:  
\begin{eqnarray}
\label{o3rugue}
 {\cal F}^B =  \frac{1}{2} \int_0^\infty dx \;    &  \Big\{ &   \Bigl   \langle \Psi_1 * \Psi_2,  W_{\frac{x}{2} + \frac{1}{2} } * B^+_L [\Psi_3 ]_{x} *[\Psi_4]_{x} * W_{\frac{x}{2} +  \frac{1}{2} } \Bigr  \rangle \\
 && +\,   (-1)^{\Psi_3} \;  \Bigl  \langle \Psi_1 * \Psi_2,  W_{\frac{x}{2} + \frac{1}{2} } * [\Psi_3 ]_{x} * B^+_L [\Psi_4]_{x} * W_{\frac{x}{2}  + \frac{1}{2} }  \Bigr \rangle \nonumber \\
 && + \,  (-1)^{\Psi_3+ \Psi_4}  \; \Bigl \langle \Psi_1 * \Psi_2,  W_{\frac{x}{2}  + \frac{1}{2} } * [\Psi_3 ]_{x} * [\Psi_4]_{x} * B^+_L W_{\frac{x}{2} + \frac{1}{2} } \Bigr \rangle \;   \Big\} \, . \nonumber
\end{eqnarray}
The first term in this amplitude is illustrated in the string diagram 
of Figure~\ref{rz6fig}. 
Restricting to the case of tachyons,
we find
\be \label{FBintegral}
{\cal F}^B =\frac{1}{2} \int_0^\infty dx \;   \left(\frac{2}{\pi}\right)^{\alpha' \sum p_i^2 -4}  \, x^{\alpha' (p_3^2 + p_4^2) -2 }  \, A^g A^m  \, ,
\ee
where the matter and ghost correlators are defined by
\begin{eqnarray}
A^m  & = & \langle  e^{i p_1 \cdot X(r_1)}    e^{i p_2 \cdot X(r_2)}   e^{i p_3 \cdot X(r_3)} e^{i p_4 \cdot X(r_4)}     \rangle_{{\cal C}_\ell} \, \\
A^g & = & - \langle  c(r_1) c(r_2) {\cal B} c(r_3) c(r_4) \rangle_{{\cal C}_\ell} +  \langle  c(r_1) c(r_2)  c(r_3) {\cal B}  c(r_4) \rangle_{{\cal C}_\ell} -  \langle  c(r_1) c(r_2)  c(r_3) c(r_4) {\cal B}\rangle_{{\cal C}_\ell} \,. \nonumber
\end{eqnarray}
Here the insertion points and circumference of the cylinder are
\be
r_1 = 0 \, , \quad r_2 = 1 \, ,\quad  r_3 = 2+ x \, , \quad  r_4 = 2 + 2 x \, , \quad \ell = 3 + 3 x \,.
\ee
We also define
\be
\gamma' = \frac{\pi}{\ell} = {\pi\over 3} \, {1\over 1+ x} \,.
\ee
The modular parameter for this geometry is
\be
\label{mod_diag3}
\lambda = \frac{\sin(\gamma') \sin(\gamma' x)}{\sin^2(\gamma'(x+2))} \,, \qquad
1-\lambda = {3\over 4} \, {1\over \sin^2(\gamma'(x+2))}\,.  
\ee
One readily finds that $\lambda (1/x) = \lambda (x)$.
The modular region $\lambda \in [0, \frac{1}{4}]$ is covered
once as $x$ goes from $0$ to $1$, and once again as $x$ goes from $1$ to $\infty$.

Using (\ref{matter4pt}), the matter correlator is
\be
\begin{split}
A^m   & =   (2 \pi)^{D} \delta \left(\sum p \right)  \, {\gamma'}^{\alpha' \sum p_i^2} \\
&\quad \cdot\, \sin(\gamma')^{-\alpha' (\s+ p_1^2 + p_2^2)} 
\cdot  \sin(\gamma' x)^{-\alpha'(\s + p_3^2 + p_4^2 )}  \\
& \quad\cdot  \sin(\gamma'(2+x))^{\alpha'(2\s + 2\t + \sum p_i^2)}
(\sqrt{3}/2)^{-\alpha'(2 \t + \sum p_i^2)}
\end{split}
\ee
From (\ref{Bccc}) we find the ghost correlator
\be
A^g
= -\frac{81}{4 \pi^3} (1+x)^2 \sin \gamma' (1-x) \, .
\ee
Assembling partial results,
\be \label{FBassemb}
\begin{split}
{\cal F}^B  & =    -(2 \pi)^{D} \delta \left(\sum p \right)    \Bigl(\frac{2}{\pi}\Bigr)^{\alpha' \sum p_i^2 -4} \int_0^\infty dx  \,{\pi\over 8}\,
 \frac{1}{(1+x)^2} ~
 {\sin \gamma' (1-x)\over \sin^4 (\gamma'(2+x))} \, \\[1.0ex]
&~~~~~~~~~\cdot  \lambda^{-\alpha'\s -2} \, (1-\lambda)^{-\alpha' \t - 2}  \, 
\Bigl({2\over \sqrt{3}} \sin(\gamma'(2+x))\Bigr)^{-4 + \alpha'\sum p^2}\cdot  \Bigl({\sin\gamma'\over \gamma'}\Bigr)^{\chi_{12}}  
\Bigl(  \frac{\sin\gamma' x}{\gamma' x}\Bigr)^{\chi_{34}} \,. 
\end{split}
\ee
Under the change of variables $x \to 1/x$, the integral (\ref{FBassemb}) goes into minus the same expression with
$\chi_{12} \leftrightarrow \chi_{34} $. Thus we can restrict the integration region to $x \in [0,1]$,
provided we antisymmetrize the integrand under the exchange 
$ \chi_{12} \leftrightarrow \chi_{34}$: 
\be
\begin{split}
{\cal F}^B  & =    -(2 \pi)^{D} \delta \left(\sum p \right)
 \Bigl(\frac{2}{\pi}\Bigr)^{\alpha' \sum p_i^2 -4} \int_0^1 dx  \,{\pi\over 8}\,
 \frac{1}{(1+x)^2} ~
 {\sin \gamma' (1-x)\over \sin^4 (\gamma'(2+x))} \, \\[1.0ex]
&~~~~~~~~~\cdot  \lambda^{-\alpha' \s -2} \, (1-\lambda)^{-\alpha' \t - 2}  \, 
\Bigl({2\over \sqrt{3}} \sin(\gamma'(2+x))\Bigr)^{-\chi} 
\\[0.5ex]
&\qquad~~ \cdot   \Bigl[  \Bigl({\sin\gamma'\over \gamma'}\Bigr)^{\chi_{12}}  
\Bigl(  \frac{\sin\gamma' x}{\gamma' x}\Bigr)^{\chi_{34}} -
\Bigl(  \frac{\sin\gamma' x}{\gamma' x}\Bigr)^{\chi_{12}} 
\Bigl({\sin\gamma'\over \gamma'}\Bigr)^{\chi_{34}}    \Bigr]\,. 
\end{split}
\ee
To change variables of integration from $x$ to $\lambda$ we use
\be
d\lambda = {\pi\over 4} {dx\over (1+x)^2}  {\sin \gamma' (1-x)
\over \sin^4 (\gamma' (2+x))} \,.
\ee
The amplitude then becomes
\be   \label{FBfinal}
\boxed{\phantom{\Biggl(}
\begin{split}
{\cal F}^B  & =    -(2 \pi)^{D} \delta \left(\sum p \right)   \int_0^{1/4} \hskip-4pt d\lambda~
 \lambda^{-\alpha' \s -2} \, (1-\lambda)^{-\alpha' \t - 2}  
\cdot \Bigl( {\pi\over 2}(1-\lambda)^{1/2}\Bigr)^{\chi}~\\[1.0ex]
&~~~~\qquad ~~~~~\qquad ~~~~~\cdot  
\frac{1}{2} \Bigl[  \Bigl({\sin\gamma'\over \gamma'}\Bigr)^{\chi_{12}}  
\Bigl(  \frac{\sin\gamma' x}{\gamma' x}\Bigr)^{\chi_{34}} -
\Bigl(  \frac{\sin\gamma' x}{\gamma' x}\Bigr)^{\chi_{12}} 
\Bigl({\sin\gamma'\over  \gamma'}\Bigr)^{\chi_{34}}    \Bigr]\,. 
\end{split}  \phantom{\Biggl(}}
\ee
The boundary term is antisymmetric under the exchange symmetry,
\be \label{antisymmetry}
\boxed{\phantom{\Biggl(}
{\cal F}^B \leftrightarrow - {\cal F}^B  \qquad {\rm under} \; \; \;\;
(p_1,p_2) \leftrightarrow (p_3,p_4) \,.
  \phantom{\Biggl(}}
\ee
In particular, when the momenta are on-shell 
$\chi_{12} = \chi_{34} = 0$ and ${\cal F}^B = 0$.

As a first check that the boundary term restores the exchange symmetry, let us extract the pole at $\alpha' \s +1 = 0$, which arises for $\lambda \sim 0$ or $x \sim 0$.
As $x\sim 0$, we have ${\sin\gamma'\over \gamma'}
\sim {2\over \pi}\,{3\sqrt{3}\over 4}$ so we get: 
\be  
{\cal F}^B |_{pole}  \simeq~\,  (2 \pi)^{D} \delta \left(\sum p \right)  
 \Bigl[- {1\over \alpha'\s +1}\Bigr]\cdot \frac{1}{2}
  \Bigl[ - \Bigl( {3\sqrt{3}\over 4}\Bigr)^{\chi_{12}}  
\Bigl(  \frac{\pi}{2}\Bigr)^{\chi_{34}} +
\Bigl(  \frac{\pi}{2}\Bigr)^{\chi_{12}} 
\Bigl( {3\sqrt{3}\over 4}\Bigr)^{\chi_{34}}    \Bigr]\,. 
\ee
Combining this with (\ref{first_term_pole_amp_final}), we find that for the full amplitude,
\be 
{\cal F}_\s |_{pole}  \simeq~  \,  (2 \pi)^{D} \delta \left(\sum p \right) \Bigl[ -{1\over \alpha'\s +1}\Bigr]
\cdot  \frac{1}{2}   \Bigl[  \Bigl( {3\sqrt{3}\over 4}\Bigr)^{\chi_{12}}  
\Bigl(  \frac{\pi}{2}\Bigr)^{\chi_{34}} +
\Bigl(  \frac{\pi}{2}\Bigr)^{\chi_{12}} 
\Bigl( {3\sqrt{3}\over 4}\Bigr)^{\chi_{34}}    \Bigr]\, .
\ee
So  the exchange symmetry holds near the pole.  
 In the next subsection we prove that
the symmetry is exactly  obeyed  for arbitrary values of the momenta.

Though symmetric,  the amplitude near the pole does  {\it not} factorize into the product
of two off-shell vertices. A factorized answer would read  
\be
{\cal F}_\s^{factor}  |_{pole}  \simeq~  \,  (2 \pi)^{D} \delta \left(\sum p \right) \Bigl[ -{1\over \alpha'\s +1}\Bigr]
\cdot     \Bigl( {3\sqrt{3}\over 4}\Bigr)^{\chi_{12}}  
\Bigl( {3\sqrt{3}\over 4}\Bigr)^{\chi_{34}}  \, .
\ee
Indeed, the coefficient $(3 \sqrt{3}/4)^{\alpha' p_i^2 -1}$ is the conformal factor that arises in inserting a tachyon vertex operator on the Witten vertex. Factorization holds in Siegel gauge, where it has a transparent geometric interpretation. The Siegel gauge  propagator is a strip of canonical width $\pi$ and length equal to the Schwinger
parameter $T$. The pole arises for $ T \to \infty$; where  
the string diagram manifestly splits into two off-shell vertices 
attached to each side of a long propagator.  By contrast, the geometric interpretation of the Schnabl propagator
does not suggest off-shell factorization.

A natural question is whether the prescription (\ref{symm_prescr}) is unique. There are other ways
to achieve a BPZ symmetric propagator, perhaps the simplest being
\be \label{alt_prescr}
{\cal P}  ~\overset{?}{=}    \lim_{\Lambda \to \infty}   {\cal P}_{\Lambda \Lambda} \,.
\ee
Using this prescription, the $x$ integration in (\ref{FBregfinal}) would range from $1$ to $\infty$. Performing the
same steps as above, we would arrive at the boundary term
\be \label{alt_boundary}
\begin{split}
{\widetilde {\cal F} } ^B  & =    -(2 \pi)^{D} \delta \left(\sum p \right)   \int_0^{1/4} \hskip-4pt d\lambda~
 \lambda^{-\alpha'\s -2} \, (1-\lambda)^{-\alpha' \t - 2}  
\cdot \Bigl( {\pi\over 2}(1-\lambda)^{1/2}\Bigr)^{\chi}~\\[1.0ex]
&~~~~\qquad ~~~~~\qquad ~~~~~\cdot  
\Bigl[  -
\Bigl(  \frac{\sin\gamma' x}{\gamma' x}\Bigr)^{\chi_{12}} 
\Bigl({\sin\gamma'\over  \gamma'}\Bigr)^{\chi_{34}}    \Bigr]\,. 
\end{split} 
\ee
The difference ${\widetilde {\cal F} }^B - {\cal F}^B$ is symmetric under  $\chi_{12} \leftrightarrow \chi_{34}$.
It follows that the total amplitude $\widetilde {\cal F}_\s$ computed  with the prescription (\ref{alt_prescr})
is symmetric just like the total amplitude ${\cal F}_\s$ computed with the prescription (\ref{symm_prescr}).
 However,  $\widetilde {\cal F}_\s$ does not agree on shell with the standard Veneziano formula,
   since  the boundary term (\ref{alt_boundary})
does  not vanish on-shell. 
As we will discuss in \S\ref{prop_revisited},  (\ref{alt_prescr}) does not really invert
the kinetic operator $\mathcal{K}$ discussed in \S\ref{derprop} while (\ref{symm_prescr})
does.  It appears that (\ref{symm_prescr}) is the only prescription leading to a four-point amplitude that is correct on-shell and has the right exchange symmetry.

\subsection{Proof of exchange symmetry}

Our final result for the off-shell Veneziano amplitude is
\be \label{finalresult}
{\cal F}_\s = {\cal F}^{(1)} +  {\cal F}^{(2)} +  {\cal F}^{(2')} +  {\cal F}^{B} \, ,
\ee
where ${\cal F}^{(1)}$, $  {\cal F}^{(2)} +  {\cal F}^{(2')}$, and ${\cal F}^{B}$ are given
in (\ref{first_term_ven_amp_v2}), (\ref{second_part_amplitude999}) and (\ref{FBfinal}), respectively. Collecting our results we have
\be  
\label{fbold_put_together}
\begin{split}
{\cal F}_\s ~=~&(2\pi)^D  \delta \bigl( \sum p\bigr)   \int_0^{{1\over 2}} 
 d\lambda\, \lambda^{-\alpha'\s -2} 
(1-\lambda)^{-\alpha' \t -2} \,\,\\
&\cdot \Bigl[ {\pi\over 2} (1- \lambda)^{1/2} \Bigr]^{\chi} \Biggl\{~ \Bigl[  { 3+ t\over \pi} \sin \bigl[ {\pi\over 3+t} \bigr] \Bigr]^{\chi}
\Bigl[ {\lambda \over t\,(1-\lambda)}\Bigr]^{\chi_{34}}
 - \int du  \Bigl| {\sin \gamma t_2\over \gamma t_2} \Bigr|^{\chi_{12}}
 {\,du^{\chi_{34}} \over du}\\[0.5ex]
&~~\hskip32pt -{1\over 2}\,  \theta\left(\frac{1}{4}-\lambda \right)
\Bigl( \Bigl| {\sin \gamma' \over \gamma' }\Bigr|^{\chi_{12}} 
  \Bigl| {\sin \gamma' x \over \gamma' x }\Bigr|^{\chi_{34}} -
   \Bigl| {\sin \gamma' \over \gamma' }\Bigr|^{\chi_{34}} 
  \Bigl| {\sin \gamma' x \over \gamma' x }\Bigr|^{\chi_{12}}\Bigr) \Biggr\}~\,.~
\end{split}
\ee
Here $\gamma = \pi/(2+ t_1 + t_2)$ and $\gamma' = \pi/(3 (1+x))$.  
The function
$\theta(1/4 - \lambda)$  in the last term is the step-function $\theta(\mu) =0$ for $\mu  < 0$, $\theta(\mu) =1$ for $\mu >0$,
and it encodes the vanishing of the boundary integrand for $\lambda > 1/4$.  The limits
of integration for $u$ are also $\lambda$ dependent, as we will discuss shortly.

We now demonstrate that ${\cal F}_\s$,  as given in (\ref{fbold_put_together}), is
symmetric under the exchange  $(p_1, p_2)\leftrightarrow (p_3, p_4)$.
Since $\s$ and $\t$ are invariant under this exchange  
and the momenta enter
$\mathcal{F}_\s$ only through $\chi_{12}$ and $\chi_{34}$,  we are 
effectively testing the symmetry of 
$\mathcal{F}_\s$ under
the exchange $\chi_{12} \leftrightarrow \chi_{34}$. We
will  show that the symmetry holds {\it locally} on moduli space,  
that is, 
before performing the $\lambda$ integration.  For this 
only the terms inside the braces $\{ \ldots \}$ need to be looked at.

\begin{figure}
\centerline{\hbox{\epsfig{figure=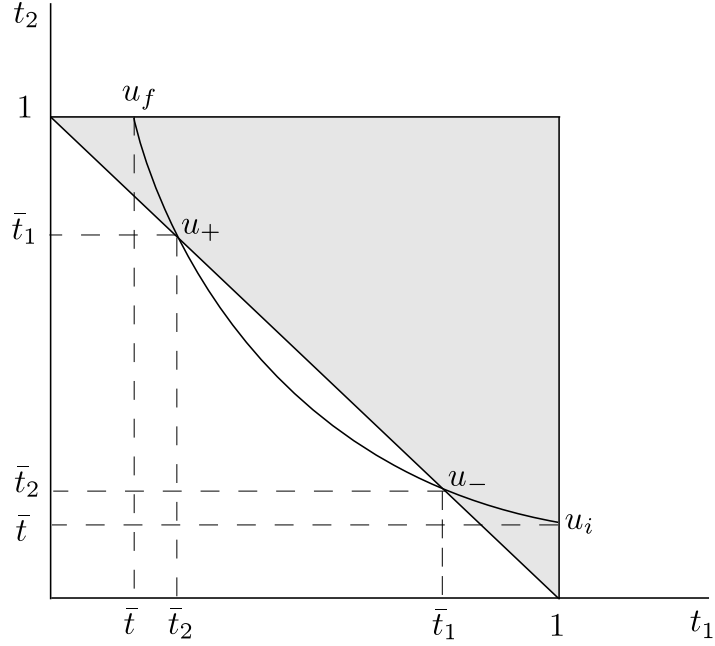, height=8.5cm}}}
\caption{Auxiliary diagram for the proof of symmetry.}
\label{rz8fig}
\end{figure}

The symmetry property would be established if 
 \be
\label{lgerkj}
\begin{split}
0=&~~\Bigl[{ \sin \bigl[ {\pi\over 3+t} \bigr]\over {\pi\over 3+ t}} \Bigr]^{\chi}
\Bigl( \Bigl[ {\lambda \over t\,(1-\lambda)}\Bigr]^{\chi_{34}}-
 \Bigl[ {\lambda \over t\,(1-\lambda)}\Bigr]^{\chi_{12}}\Bigr)\\[1.0ex]
 & - \int du\; \Bigl(  \Bigl| {\sin \gamma t_2\over \gamma t_2} \Bigr|^{\chi_{12}}
 {\,du^{\chi_{34}} \over du} 
 - \Bigl| {\sin \gamma t_2\over \gamma t_2} \Bigr|^{\chi_{34}}
 {\,du^{\chi_{12}} \over du}  \Bigr) \\[1.0ex]
 &-  \theta\left(\frac{1}{4}-\lambda \right)\Bigl( \Bigl| {\sin \gamma' \over \gamma' }\Bigr|^{\chi_{12}} 
  \Bigl| {\sin \gamma' x \over \gamma' x }\Bigr|^{\chi_{34}} -
   \Bigl| {\sin \gamma' \over \gamma' }\Bigr|^{\chi_{34}} 
  \Bigl| {\sin \gamma' x \over \gamma' x }\Bigr|^{\chi_{12}}\Bigr)\,.
\end{split}
\ee
We begin by showing that the middle term 
is the integral of a total derivative.     We recall the definition of $u$ and introduce
a companion variable $v$:
\be
u(t_1,t_2)=  {\sin \gamma t_1\over \gamma t_1}\,, \qquad
v(t_1, t_2) =   {\sin \gamma t_2\over \gamma t_2} \,.
\ee
Consider one of the curves of constant $\lambda$
in the $t_1, t_2$ diagram (Figure~\ref{rz8fig}).  
The curve is 
invariant under the reflection $t_1 \leftrightarrow t_2$  
and goes from $u= u_i$ to $u= u_f$.  Consider
a parameterization of this curve with a parameter $ \sigma \in [0,1]$: 
\be
t_1 = t_1(\sigma) \,,   ~~  t_2 = t_2 (\sigma) \,,
\ee
with the condition that the points at $\sigma$ and at $1-\sigma$ are reflections
of one another:
\be
t_1 (1-\sigma) = t_2(\sigma)  \,, ~~  t_2 (1-\sigma) = t_1 (\sigma) \,.
\ee
Through these relations we can now view $u$ and $v$, on the constant
$\lambda$ curve as just functions of $\sigma$: $u(\sigma)$, $v(\sigma)$.  We demand that  
$u(0) = u_i$ and $u(1) = u_f$.   Given the parameterization and the fact that
$u \leftrightarrow v$ as $t_1 \leftrightarrow t_2$,  we have
\be
v(\sigma) = u (1-\sigma)\,.
\ee 
The middle term in  (\ref{lgerkj}) is
\be
\begin{split}
&~~~ - \int_{u_i}^{u_f} d \sigma \Bigl( v(\sigma)^{\chi_{12}}
 {\,d \over d \sigma} u( \sigma)^{\chi_{34}}
 - v( \sigma)^{\chi_{34}}
 {\,d \over d \sigma} u( \sigma)^{\chi_{12}} \Bigr)\\[1.0ex]
 &=   - \int_0^1 d \sigma \Bigl( u(1- \sigma)^{\chi_{12}}
 {\,d \over d \sigma} u( \sigma)^{\chi_{34}}
 - u(1- \sigma)^{\chi_{34}}
 {\,d \over d \sigma} u( \sigma)^{\chi_{12}} \Bigr)\\[1.0ex]
 &=   - \int_0^1 d \sigma \Bigl( u(1-\sigma)^{\chi_{12}}
 {\,d \over d \sigma} u( \sigma)^{\chi_{34}}
 +  {\,d \over d \sigma} u(1- \sigma)^{\chi_{12}}u( \sigma)^{\chi_{34}} \Bigr)\\[1.0ex]
 &=   - \int_0^1 d \sigma {d\over d \sigma } \Bigl( u(1- \sigma)^{\chi_{12}}
 u( \sigma)^{\chi_{34}} \Bigr)\\[1.0ex]
 &=   - \Bigl( u(0)^{\chi_{12}}
 u(1)^{\chi_{34}} -  u(1)^{\chi_{12}}
 u(0)^{\chi_{34}}  \Bigr)
 = ~  u_f^{\chi_{12}}
 u_i^{\chi_{34}} -  u_f^{\chi_{34}} u_i^{\chi_{12}}\,.
 \end{split}
\ee
In the last equality we assumed $\lambda > 1/4$, in which case the 
constant $\lambda$ curves
that do not intersect the diagonal and  the integration region indeed
goes from $u_i$ to $u_f$.
For $\lambda < 1/4$ the integral is really of the form
\be
- \int_{u_i}^{u_-}\ldots  - \int_{u_+}^{u_f} \ldots  =  - \int_{u_i}^{u_f}\ldots
+  \int_{u_-}^{u_+}\ldots
\ee
where $u_- < u_+$ are the points on the curve that are also
on  the diagonal $t_1 + t_2= 1$ (Figure~~\ref{rz8fig}).  Therefore, for $\lambda < 1/4$ we get the {\em extra} terms
\be
-u_+^{\chi_{12}}
 u_-^{\chi_{34}} +  u_+^{\chi_{34}} u_-^{\chi_{12}}\,.
\ee
It now follows that  (\ref{lgerkj})  holds if  the following two identities
are satisfied:
\be
\label{lgerkj1}
0=\Bigl[{ \sin \bigl[ {\pi\over 3+t} \bigr]\over {\pi\over 3+ t}} \Bigr]^{\chi}
\Bigl( \Bigl[ {\lambda \over t\,(1-\lambda)}\Bigr]^{\chi_{34}}-
 \Bigl[ {\lambda \over t\,(1-\lambda)}\Bigr]^{\chi_{12}}\Bigr)
 +  u_f^{\chi_{12}}
 u_i^{\chi_{34}} -  u_f^{\chi_{34}} u_i^{\chi_{12}}
 \ee
\be
\label{lgerkj2}
0= -u_+^{\chi_{12}}
 u_-^{\chi_{34}} +  u_+^{\chi_{34}} u_-^{\chi_{12}}
 -\Bigl( \Bigl| {\sin \gamma' \over \gamma' }\Bigr|^{\chi_{12}} 
  \Bigl| {\sin \gamma' x \over \gamma' x }\Bigr|^{\chi_{34}} -
   \Bigl| {\sin \gamma' \over \gamma' }\Bigr|^{\chi_{34}} 
  \Bigl| {\sin \gamma' x \over \gamma' x }\Bigr|^{\chi_{12}}\Bigr) \,.
\ee
The above identities are antisymmetric under the exchange $\chi_{12} \leftrightarrow \chi_{34}$.
It is then sufficient to  show that:
\be
\label{lgerkj11}
\begin{split}
0&=\Bigl[{ \sin \bigl[ {\pi\over 3+t} \bigr]\over {\pi\over 3+ t}} \Bigr]^{\chi} \Bigl[ {\lambda \over t\,(1-\lambda)}\Bigr]^{\chi_{34}}  -  u_f^{\chi_{34}} u_i^{\chi_{12}} \, ,\\[1.0ex]
 0&= -u_+^{\chi_{12}}
 u_-^{\chi_{34}} 
+   \Bigl| {\sin \gamma' \over \gamma' }\Bigr|^{\chi_{34}} 
  \Bigl| {\sin \gamma' x \over \gamma' x }\Bigr|^{\chi_{12}} \,.
  \end{split}
\ee

Consider the first relation.  It follows from the modulus calculation
in the first diagram that 
\be
 {\lambda \over t\,(1-\lambda)} = { \sin \bigl[ {\pi t\over 3+t} \bigr]\over 
 t \sin\bigl[ {\pi\over 3+ t}\bigr]} \,.
\ee
Then
\be
\label{id_3}
\Bigl[{ \sin \bigl[ {\pi\over 3+t} \bigr]\over {\pi\over 3+ t}} \Bigr]^{\chi} \Bigl[ {\lambda \over t\,(1-\lambda)}\Bigr]^{\chi_{34}}  =
 \Biggl[ { \sin \bigl[ {\pi  t\over 3+ t} \bigr]\over 
 {\pi t\over 3+t}} \Biggr]^{\chi_{34}} \Biggl[
{ \sin \bigl[ {\pi \over 3+ t} \bigr]\over 
  {\pi\over 3+  t}}\Biggr]^{\chi_{12}} \,.
 \ee
The second term on the first relation is associated with the second diagram.
The point $u_i$ corresponds to $t_1=1$ and some value $t_2 = \bar t$.
The point $u_f$ corresponds to $t_2=1$ and $t_1 = \bar t$ (see Figure~\ref{rz8fig}). Hence,
\be
\label{id_12}
u_f =
  { \sin \bigl[ {\pi \bar t\over 3+\bar t} \bigr]\over 
 {\pi\bar t\over 3+\bar t}}\,,~~
 u_i =  { \sin \bigl[ {\pi \over 3+\bar t} \bigr]\over 
  {\pi\over 3+ \bar t}}\,.
\ee
The modulus $\lambda$ for the curve in question is given by (\ref{mod_diag2}),
using $t_1=1, t_2 = \bar t$:
\be
\label{mod_diag222}
\lambda  
=  \frac{\sin\bigl[{\pi \over 3+ \bar t} \bigr] \sin\bigl[{\pi \bar t \over 3+ \bar t} \bigr] }{ \sin^2\bigl[{2\pi \over 3+ \bar t} \bigr] } \,.
\ee
Comparing with (\ref{mod_diag1}) we see that for fixed $\lambda$ we have $\bar t = t$.
It then follows from (\ref{id_12}) that  $u_f^{\chi_{34}}
 u_i^{\chi_{12}} $ coincides with the right hand side of (\ref{id_3}).
This proves the first equality in (\ref{lgerkj11}).

To prove the second relation in (\ref{lgerkj11}) we first note that $u_-$ 
corresponds to $t_1 + t_2 = 1$, which gives $\gamma = \pi/3$, and 
some values $\bar t_1$ and $\bar t_2$ for $t_1$ and $t_2$, such that $\bar t_1 >1/2$ (see Figure 10).
The modulus associated with $u_-$ (or $u_+$, since it lies on the same
curve) is  (\ref{mod_diag22}),
\be
\label{mod_diag22xx}
1-\lambda = ~ {3\over 4}\, {1\over \sin^2({\pi\over 3} (\bar t_1+1))} \,.
\ee
For the second term on the right-hand side of the 
equality the modulus is the function of $x$ given by
(\ref{mod_diag3}),
\be
\label{mod_diag3x}
1-\lambda = {3\over 4} \, {1\over \sin^2 \bigl[  {\pi (x+2)\over 3(1+x)}\bigr]} \,.
\ee
It follows from the last two equations that 
\be  \bar t_1 = {1\over 1+ x} \quad \to \quad  \gamma' = {\pi\over 3} {1\over 1+ x} = \gamma \bar t_1 \,, ~~\gamma' x = \gamma \bar t_2\,.
\ee
Since $0 < x < 1$, we have $\bar t_1 < 1/2$, as required.  
Therefore:
\be
{\sin \gamma'\over \gamma'} =  {\sin \gamma \bar t_1\over \gamma \bar t_1} = u_-\,,~~\hbox{and}~~
{\sin \gamma' x\over \gamma' x} =  {\sin \gamma \bar t_2\over \gamma \bar t_2} = u_+\,.
\ee
These relations make it clear that the second equation in (\ref{lgerkj11}) also holds.
This completes  the proof that $\mathcal{F}_\s$  
has the requisite  exchange symmetry.

Having reassured ourselves that the symmetry holds, we can write a {\it manifestly}
symmetric form of the amplitude by averaging over the two orderings 1234 and 3412,
\be
\label{Final}
\boxed{\phantom{\Biggl(}
\begin{split}
{\cal F}_\s ~=~&(2\pi)^D  \delta \bigl( \sum p\bigr)   \int_0^{{1\over 2}} 
 d\lambda\, \lambda^{-\alpha'\s -2} 
(1-\lambda)^{-\alpha' \t -2} \cdot \Bigl[ {\pi\over 2} (1- \lambda)^{1/2} \Bigr]^{\chi}
\,\,\\
& \Biggl\{~ 
\frac{1}{2} \Bigl[  { 3+ t\over \pi} \sin \bigl[ {\pi\over 3+t} \bigr] \Bigr]^{\chi} 
\left( 
\Bigl[ {\lambda \over t\,(1-\lambda)}\Bigr]^{\chi_{34}} + \Bigl[ {\lambda \over t\,(1-\lambda)}\Bigr]^{\chi_{12}} \right)
\\[1.0ex] 
&~~ -  \frac{1}{2} \int du \;  \; \left(  \Bigl| {\sin \gamma t_2\over \gamma t_2} \Bigr|^{\chi_{12}}
 {\,du^{\chi_{34}} \over du}   +     \Bigl| {\sin \gamma t_2\over \gamma t_2} \Bigr|^{\chi_{34}}
 {\,du^{\chi_{12}} \over du}  \right)   \Biggr\} \,. \\
\end{split}}
\ee
This is the final and most useful form of  ${\cal F}_\s$.  For ease of reference, we
recall that the variable $t$ is a
 function of the modulus $\lambda$, given by (\ref{toflambda}); 
 the variable $t_2$ should 
 be thought as a function of $\lambda$ and of the integration variable $u$,
 according to the definitions given in (\ref{mod_diag2}) and 
 (\ref{udef});
the variables $\chi$, $\chi_{12}$ and $\chi_{34}$ are functions of the external momenta,
given in (\ref{chidef}).  The limits of integration for $u$ are from $u_i$ to $u_f$
for $1/4\leq \lambda \leq 1/2$ and from $u_i$ to $u_-$ together with $u_+$ to $u_f$,
for $0\leq \lambda \leq 1/4$.

\medskip
Our computation has given us the full $\s$-channel contribution $\mathcal{F}_\s$ 
to the amplitude when the punctures appear as $1234$ as we travel on the boundary of the circle in the
counterclockwise direction.  We can rewrite
(\ref{Final})  as 
\be
{\cal F}_\s ~=~(2\pi)^D  \delta \bigl( \sum p\bigr)   \int_0^{{1\over 2}} 
 d\lambda \,\lambda^{-\alpha'\s -2} (1-\lambda)^{-\alpha' \t -2}  \, F (\lambda , \chi_{ij})\,,
\ee
where introduced the function $F$ of the modulus $\lambda$ and of the $\chi_{ij}$ invariants.
As we discussed at the end of \S\ref{firstdiagram}, the $\t$-channel contribution would be
given by 
\be
{\cal F}_\t ~=~(2\pi)^D  \delta \bigl( \sum p\bigr)   \int_{1/2}^{{1}} 
 d\lambda \,\lambda^{-\alpha'\s -2} (1-\lambda)^{-\alpha' \t -2}  \, 
 F (1-\lambda , \chi_{i+1,j+1})\,.
\ee
The off-shell continuity of the integrand for combined $\s$- and $\t$-channel 
amplitudes requires $F(1/2, \chi_{ij}) = 
F(1/2, \chi_{i+1,j+1})$.  This holds on account of our earlier analysis, since
neither the second diagram nor the third one contribute for $\lambda = 1/2$.
The off-shell Veneziano amplitude $A(\s,\t)$ is  
given by
\be
\label{osvam}
A(\s,\t) = {\cal F}_\s + {\cal F}_\t \,.
\ee
The full four-tachyon amplitude is obtained by adding to $A(\s,\t)$
the other terms that correspond to in-equivalent orderings of 
four punctures on the boundary of a disk.

\subsection{The propagator revisited}\label{prop_revisited}

We have defined a propagator $\mathcal{P}$ in (\ref{symm_prescr})
that leads to a sensible off-shell amplitude for four tachyons.  It is natural
now to revisit the formal computations of \S\ref{derprop} to see if this
propagator, with its regulation and symmetrized limits, provides an inverse
to the kinetic operator on the gauge slice.  We ask if the propagator
$\mathcal{P}$ really satisfies $\mathcal{P} \mathcal{K} = P$ where, as before,
$P= BC$ is the projector to the gauge slice.  As it turns
out this equation is {\em not} satisfied.  What we will find is that, with our definition,
\be
\label{prop_goal}
\mathcal{P} \mathcal{K} = P  +  Q \eta \,.
\ee 
In the above operator equation we find  $BQ\eta=0$ and $B\eta=0$.  
Acting on arbitrary states, the extra term on the 
right-hand side is therefore a truly trivial state in the gauge slice.  If a gauge
condition fixes the gauge completely such states cannot exist because
they could be added to any physical state while preserving the gauge condition.
It may be that the appearance of the extra term indicates that 
Schnabl gauge has additional subtleties.  
With a unitary matter CFT,  Siegel gauge has no trivial states on the gauge slice. 
Nor there are such states in 
Siegel gauge at zero momentum in the standard flat spacetime background. 
In Schnabl gauge, however, there
is a  BRST trivial state $\psi_0'$ that satisfies the gauge condition, making it impossible
to fix the gauge completely.  The state $\psi_0' \sim Q BL^+ c_1|0\rangle$ 
indeed satisfies the gauge
condition: $BQ BL^+ c_1|0\rangle =  LBL^+ c_1|0\rangle = BLL^+ c_1|0\rangle =0$.
It is
truly trivial since it is $Q$ of something which is in the gauge.
The state $\psi_0'$ plays a key role in the tachyon vacuum solution and
features in the difficulties to construct exactly marginal
solutions for general operators in Schnabl gauge~\cite{KORZ}.   

If we do not use the symmetrized prescription for $\mathcal{P}$ we cannot
even obtain (\ref{prop_goal}).  We will not attempt here to discuss the possible
implications of (\ref{prop_goal}) for the computation of string amplitudes, nor if
it reflects a shortcoming of the gauge condition or a shortcoming of the presently defined
propagator.

To derive (\ref{prop_goal}) we begin with 
\be
\mathcal{P}_{\Lambda \Lambda^\star} \mathcal{K}  =
{B\over L_\Lambda} Q {B^\star\over L^\star_{\Lambda^\star}} \bigl( - C^\star
B^\star Q B C ) = {B\over L_\Lambda}  {L^\star\over L^\star_{\Lambda^\star}}  Q B C 
= {B\over L_\Lambda}  \bigl( 1 - e^{-\Lambda^\star L^\star}\bigr)  Q B C \,,
\ee
and rewrite this as 
\be
\begin{split}
\mathcal{P}_{\Lambda \Lambda^\star} \mathcal{K} & =
{B\over L_\Lambda} Q B C - {B\over L_\Lambda} Q\,  e^{-\Lambda^\star L^\star} B C\, 
\\
 & =
{L\over L_\Lambda}  B C - {L\over L_\Lambda}   e^{-\Lambda^\star L^\star} B C
+  Q \,B\,{1\over L_\Lambda}   e^{-\Lambda^\star L^\star} B C\,.
\end{split}
\ee
  Expanding the first two terms one
gets
\be
\label{get_prop_this}
\mathcal{P}_{\Lambda \Lambda^\star} \mathcal{K}  =\bigl(
1- e^{-\Lambda L} -e^{-\Lambda^\star L^\star}
+ e^{-\Lambda L} e^{-\Lambda^\star L^\star} \bigr) BC 
+  Q B\, {1\over L_\Lambda}   e^{-\Lambda^\star L^\star}B C\,.
\ee
Consider  the terms inside the parentheses.
As $\Lambda$ becomes large the term $e^{-\Lambda L}$ will give a singular
surface -- thus no contribution. The same holds for $e^{-\Lambda^\star L^\star}$
for large $\Lambda^\star$.  The last term, $e^{-\Lambda L} e^{-\Lambda^\star L^\star}$,
is more delicate.  It vanishes in our prescription, since one gets a singular surface
as soon as one parameter ($\Lambda$ or $\Lambda^\star$) goes to infinity.
Interestingly, if we take $\Lambda = \Lambda^\star$ then 
$e^{-\Lambda L} e^{-\Lambda L^\star}\sim e^{-L^+}$ and this gives a regular
surface.  We thus see that, apart from the $Q$ exact term, we get the 
expected answer -- just $BC$ -- for the symmetrized prescription and not
from the prescription (\ref{alt_prescr}), for example.

Let us now examine the extra $Q$ trivial term in (\ref{get_prop_this}). Comparing
with (\ref{prop_goal}) we identify~$\eta$:
\be
\eta =  B\, {1\over L_\Lambda}   e^{-\Lambda^\star L^\star} BC  \,.
\ee
Clearly $B \eta =0$.  Let us now see if $BQ \eta = 0$.  Using also $B e^{-\Lambda^\star L^\star} \simeq -  e^{-\Lambda^\star L^\star} B^\star$, which holds for 
large $\Lambda^\star$ (see (\ref{belstar})), we get

\be
B Q \eta  =\{ B, Q\} \eta =   {L\over L_\Lambda} B  e^{-\Lambda^\star L^\star} BC
= - \bigl(  1 - e^{-\Lambda L} \bigr) e^{-\Lambda^\star L^\star} B^\star BC\,.
\ee
With the symmetrized prescription the final right-hand side vanishes, as desired.  
We thus confirm the structure predicted in
(\ref{prop_goal}).  This is  evidence that our definition of the propagator
is consistent with its fundamental role as an inverse of the kinetic term.

\medskip
Before concluding this section we discuss another rewriting of the propagator.
With the benefit of hindsight we can write the propagator in a form that makes the
exchange symmetry more transparent. Instead of commuting $Q$ to the right as in (\ref{moveQ}),
we {symmetrize}  over the two ways of commuting  $Q$, 
\be \label{symQP}
{\cal P}_{\Lambda \Lambda^\star}   =     \frac{1} {2}  \Bigl[\,\frac{B}{L_\Lambda}    \; - \frac{B}{L_\Lambda} \frac{B^\star}{L^\star_{\Lambda^\star}}\,  Q      
\;  -    \frac{B}{L_\Lambda}    \,e^{-\Lambda^\star L^\star} \Bigr] + \; 
 \frac{1} {2}  \Bigl[\, \frac{B^\star}{L^\star_{\Lambda^\star }}  -      
 Q \, \frac{B}{L_\Lambda }  
 \frac{B^\star}{L^\star_{\Lambda^\star}}       
\;  -   e^{-\Lambda L } \frac{B^\star}{L^\star_{\Lambda^\star   }}  \; \Bigr] \,.
\ee
The first group of terms is the result of moving $Q$ to the right
and the second group is  the result of moving $Q$ to the left. There
are  now two boundary terms, 
which we can bring to a common form by a few applications of the CBH formula. For the first boundary term,
\begin{eqnarray} \label{firstbound}
  \frac{B}{L_\Lambda}    \,e^{-\Lambda^\star L^\star}  &  = &  
    \int_{s^\star\over s }^{s^\star}  \frac{dx}{x}  \; \; e^{-x \left(1-\frac{1}{s^\star} \right) L^+} \;  \; e^{\left[ \left(x-\frac{1}{x}\right) + \frac{1-x}{s^\star}  \right] L} \cdot \Bigl[  \frac{1}{s^\star} B + \Bigl(\frac{1}{s^\star}-1 \Bigr) B^\star\Bigr]
\\[0.5ex] 
& \cong &    \int_{\frac{s^\star}{s} }^{\infty}  \frac{dx}{x} \;  \; e^{-x  L^+} \;  \; e^{ \left(x-\frac{1}{x}\right) L} \cdot  (-  B^\star )\, . \nonumber
\end{eqnarray}
 In the last line we have dropped terms that are subleading for large $s = e^\Lambda$ and $s^\star = e^{\Lambda^\star}$, irrespective of how the limits are taken.
 For the second boundary term,
\begin{eqnarray} \label{secondbound}
 e^{-\Lambda L}   \frac{B^\star}{L^\star_{\Lambda^\star } }   &  = &  
  \int^{\frac{s^\star}{s} }_{\frac{1}{s}}  \frac{dx}{x}  \; \;  e^{  - \left( x  - \frac{1}{s} \right) L^+} \; \; e^{\left[ \left(x-\frac{1}{x}\right) +  \frac{1}{s} \left(  \frac{1}{x} -1 \right)  \right] L} \, B^\star
\\ & \cong &    \int^{\frac{s^\star}{s} }_{0}  \frac{dx}{x}  \; \; e^{-x  L^+} \; \;  e^{ \left(x-\frac{1}{x}\right) L} \cdot  B^\star \, . \nonumber
\end{eqnarray}
Apart from the integration ranges, the two boundary terms are exactly equal and opposite. According to the prescription  (\ref{symm_prescr}), we are instructed
to {average} each term over the two values $s^\star/s =0$ and $s^\star/s = \infty$. 
If we take $s^\star/s = \infty$ in (\ref{firstbound})
we have a singular expression with $x \to \infty$ and, similarly, if we take $s^\star/s = 0$ in (\ref{secondbound}) we have a singular
expression with $x \to 0$. When  the propagator  is inserted in a generic string diagram, these terms correspond
to codimension one loci in moduli space (since $x$ is fixed to a specific value), where the surface degenerates. 
It seems safe to assume that these terms give a vanishing contribution. The two remaining possibilities correspond to taking $s^\star/s = 0$ in (\ref{firstbound})
and $s^\star/s = \infty$ in (\ref{secondbound}). The range of  $x$ is from 0 to $\infty$ in both cases and the two terms
cancel each other out.

In summary,  when the $Q$ is moved symmetrically to the left and the right 
the boundary terms
cancel and the propagator can be written as
\be \label{symQPnoprop}
{\cal P}_{\Lambda \Lambda^\star}    =      \frac{1} {2}  \Big[ \frac{B}{L_\Lambda}   
  - \frac{B}{L_\Lambda}    \frac{B^\star}{L^\star_{\Lambda^\star}}     Q      
 +     \frac{B^\star}{L^\star_{\Lambda^\star }}  -      Q  \frac{B}{L_\Lambda} 
 \frac{B^\star}{L^\star_{\Lambda^\star} }       
\;    \Big] \, , 
\ee
with the usual understanding that the cutoffs are removed according to (\ref{symm_prescr}). In fact, when (\ref{symQPnoprop})
is inserted in a tree level four-point amplitude,  the $\Lambda \to \infty $ and $\Lambda^\star \to \infty$ limits commute.  The precise definition (\ref{symm_prescr}) may be necessary for more complicated amplitudes.

\sectiono{Discussion}

The key new features of the Schnabl gauge propagator are the presence
of the BRST operator $Q$ and the need for two Schwinger parameters
to represent the operators $1/L$ and $1/L^\star$.  The BRST action on
differential forms on moduli space is 
familiar~\cite{AlvarezGaume:1988bg,Zwiebach:1992ie}.  Given 
some degree $k$ form $\Omega_{\Psi_1 \cdots \Psi_n}$
labeled by external states $\Psi_i$,  BRST action gives  degree $k$ forms
$\Omega_{\sum_i \Psi_1 \cdots Q\Psi_i \cdots  \Psi_n}$where $Q$ acts on the states, and 
the degree $k+1$ exterior derivative $d\Omega_{\Psi_1, \cdots, \Psi_n}$.  
Since string amplitudes
are integrals of differential forms over moduli space, the action of the various $Q$'s
arising from the various propagators results in two
effects:  
\begin{enumerate}

\item  Some $Q$'s end up acting on external 
states and, 

\item  Some $Q$'s set  Schwinger parameters to limit values.
\end{enumerate}

For any given string amplitude the diagrams 
in which $Q$'s act on external states vanish on-shell. 
These diagrams have more Schwinger parameters than those
needed to produce the moduli space, one extra parameter for each $Q$ that acts
on an external state.  Our example made this clear: the four-point amplitude
includes a string diagram (diagram two) with a  $Q$ acting on external states and,  in addition to the modular parameter $\lambda$, one extra parameter of integration $u$.

The second effect of $Q$ is more subtle. Previously it was thought
that the only relevant boundaries arise when Schwinger parameters
go to zero, namely, from collapsed propagators.  In the four-string
amplitude this familiar boundary gave the on-shell amplitude (diagram one).
We have found that a boundary at infinite value of the Schwinger
parameter -- naively a degenerate surface -- can also give a
regular contribution.   In the four-string amplitude
it gave a boundary term that vanished on-shell but 
helped restore the off-shell exchange symmetry.
We do not know if boundaries at infinite values of 
Schwinger parameters can contribute on-shell in general string amplitudes.

We have obtained  a symmetrized form (\ref{symQPnoprop}) of the propagator that
is written
without boundary terms.  In a string diagram with various propagators, boundary
terms will arise as the $Q$'s are moved across $B/L$'s or $B^\star/L^\star$'s in order
to get them to act on the external states.  We have not discussed  the rules needed to deal with these boundaries.  Thus, work remains to be done to fully 
understand tree amplitudes.  
A better understanding of the propagator might 
arise by further analysis of how it defines an inverse, following the 
preliminary discussion
at the beginning of \S\ref{prop_revisited}.  

In Siegel gauge, an open string diagram has a {\em spine}: the line formed
by the set of all string midpoints on all the propagators -- on each strip, the line
parallel to the boundaries that divides the strip in two equal parts.  In Schnabl
gauge the string diagrams map all open string midpoints to infinity.
 In the disk picture, for example, 
 propagators and states are wedges
and all string midpoints are at the center of the disk. The spine 
thus collapses to a point.  
It appears that this gives interesting complications for open string loop diagrams.
In a planar one-loop amplitude the spine is a curve homotopic to the two 
boundaries of the annulus diagram.  If the spine collapses the annulus becomes
singular -- it produces the closed string degeneration.  It is not clear how finite
modulus annuli are produced. 

It used to be thought that covariant open string field theory is 
canonically associated with a certain way to decompose the moduli
space of Riemann surfaces~\cite{Giddings:1986wp,Zwiebach:1990az} -- through Strebel quadratic 
differentials and minimal area string diagrams.  We now see that in Schnabl
gauge (and presumably in any projector gauge) one finds rather new
kind of string diagrams.  A good geometrical understanding of these
diagrams may have significant implications for closed string
field theory and for closed string physics in open string field theory.

\vspace{1cm}

{\bf \large Acknowledgments}

\medskip

We thank Yuji Okawa for useful comments on a draft version
of this paper.  LR wishes to thank the Galileo Institute
for Theoretical Physics (Florence) and  Scuola Normale Superiore (Pisa)
for their great hospitality and the INFN for partial support during the completion of this work.
The work of B.Z. is supported in part
by the U.S. DOE grant DE-FC02-94ER40818.
The work of L.R. is  supported in part
by the National Science Foundation Grant No. PHY-0354776
and by the DOE Outstanding Junior Investigator Award. 
Any opinions, findings, and conclusions or recommendations
expressed in this material are
those of the authors and do not necessarily reflect the views of
the National Science Foundation.

\medskip

\appendix

\sectiono{Notation and algebraic identities}  

In this appendix we collect our main definitions and several useful algebraic identities.
We refer to \cite{RZ, ORZ, KORZ} for a more detailed exposition of many of these facts.

\subsection{Basic properties}
The operators $L$ and $B$ are the zero modes of the stress tensor and of the antighost
in the conformal frame of the sliver,
\be
L \equiv \oint \frac{d \xi}{2 \pi i}   \, \frac{f(\xi)}{f'(\xi)} T(\xi) \, , \quad B  \equiv 
\oint \frac{d \xi}{2 \pi i}   \, \frac{f(\xi)}{f'(\xi)} 
b(\xi) \, , \quad f(\xi) \equiv \frac{2}{\pi} \arctan \xi \,.
\ee
The superscript $\star$ indicates BPZ conjugation. We define
\be
L^\pm = L \pm L^\star\, , \quad B^\pm = B \pm B^\star\, .
\ee
Two basic commutation relations are
\be \label{basic[]}
[ L, L^+ ] = L^+ \, , \qquad  [B, L^+] = B^+ \, .
\ee
The subscripts $L$ and $R$ denote the left and the right part of an operator. We have
\be
L^+ = L^+_L + L^+_R \, , \quad B^+ = B^+_L + B^+_R \, .
\ee
From (\ref{basic[]})
we deduce
\be
[ L, L^+_L ] = L^+ \, , \qquad  [B, L^+_L] = B^+_L \, .
\ee
The operators $L - L^+_L$,  $L^\star - L^+_L$, $B - B^+_L$, and 
$B^\star - B^+_L$
 are all derivations of the star algebra. This implies a simple action of $L$, $L^\star$,
 $B$, $B^\star$ on products of string fields:
\begin{eqnarray}
\label{derivation_type}
 L (\phi_1 *  \ldots * \phi_n)  & =  & (L \phi_1) *\ldots *\phi_n
+\sum_{m=2}^n  
 ~\phi_1 * \ldots  *(L-L^+_L) \phi_m  *\ldots * \phi_n \,, \\
  L^\star (\phi_1 *  \ldots * \phi_n)  & =  & (L^\star \phi_1) *\ldots *\phi_n
+\sum_{m=2}^n  
 ~\phi_1 * \ldots  *(L^\star-L^+_L) \phi_m  *\ldots * \phi_n \,, \nonumber \\
B (\phi_1 *  \ldots * \phi_n)  & = &   (B \phi_1) *\ldots *\phi_n
+\sum_{m=2}^n  (-1)^{\sum_{k=1}^{m-1}\phi_k}
 ~\phi_1 * \ldots  *(B-B^+_L) \phi_m  *\ldots  *\phi_n \,, \nonumber \\
B^\star (\phi_1 * \ldots * \phi_n)   & =  & (B^\star \phi_1) *\ldots *\phi_n
+\sum_{m=2}^n  (-1)^{\sum_{k=1}^{m-1}\phi_k}
 ~\phi_1 * \ldots  *(B^\star-B^+_L) \phi_m  *\ldots  *\phi_n \,. \nonumber
\end{eqnarray}
Here and elsewhere, a string field in the exponent of $-1$
denotes its Grassmann property: it is $0$ mod $2$
for a Grassmann-even string field and $1$ mod $2$
for a Grassmann-odd string field. 
We also have 
\be
\label{large_act}
\begin{split}
 e^{-TL}  (\phi_1 * \phi_2 *  \ldots * \phi_n) &=  e^{-TL}  \phi_1 *
 e^{-T(L- L_L^+)}\phi_2 * \ldots  e^{-T(L- L_L^+)}\phi_n\,, \\
 e^{-TL^\star}  (\phi_1 * \phi_2 *  \ldots * \phi_n) &=  e^{-TL^\star}  \phi_1 *
 e^{-T(L^\star - L_L^+)}\phi_2 * \ldots  e^{-T(L^\star- L_L^+)}\phi_n \,.
 \end{split}
 \ee

\subsection{Reordering formulas}

We often need to reorder exponentials of operators. In all cases,
the relevant  Lie algebra is two-dimensional, with generators $x$ and $y$ and commutation relation
$[\, x , y \,] = y$.  Algebraic identities are most easily derived
by using the explicit
(adjoint) representation of $x$ and $y$:
\be \label{xyadjoint}
x = \begin{pmatrix} 0 & 1 \cr 0 & 1 \end{pmatrix}\,, \quad 
y = \begin{pmatrix} -1 & 1 \cr -1 & 1 \end{pmatrix} \,.
\ee
As two by two matrices, $x^2 =x$, $xy=y$ and $yx = y^2 = 0$.
One then verifies that
\be
\label{cbh_2_dim}
e^{\alpha x + \beta y } =  e^{{\beta\over \alpha} (e^\alpha -1) y}  \, e^{\alpha x}\,, \quad
\hbox{when} \quad 
[\, x, y \,] = y \,.
\ee
Recurrent special cases are:
\begin{eqnarray}
\label{ordering1}  
e^{-T (L-L^+_L)}  & = & e^{(1-e^{-T}) L^+_L} \, e^{-T L} \, , \qquad x=L\,,~ y=L^+_L   \\
\label{ordering2}
e^{T (L-L^+_R)}  & = & e^{-(e^{T}-1) L^+_R} \, e^{T L} \,  , \;  \; \qquad x=L\, ,~ y=L^+_R \,. 
\end{eqnarray}
Other identities that can be proved by similar methods are:
\begin{eqnarray}
e^{-T L} B^+_L & = & e^{-T}  B^+_L e^{-T L} \,,
\\
e^{-T (L-L^+_L)} B^+_L  & =  &e^{-T}  B^+_L e^{-T (L-L^+_L) } \,,
\\
e^{-T L^\star} B^+_L  & =  & e^{T}  B^+_L e^{-T L^\star} \,,
\\
e^{-T (L^\star-L^+_L)} B^+_L   & =  &e^{T}  B^+_L e^{-T (L^\star-L^+_L) } \,.
\end{eqnarray}

\subsection{Wedge states}

The wedge states $W_\alpha$, $\alpha \geq 0$,
 are surface states defined by their overlap with a generic Fock state $\phi$,
\be
\langle W_\alpha\,, \, \phi \rangle \equiv  \langle f^{-1} \circ \frac{2}{1 + \alpha} \circ f \circ \phi(0) \rangle_{UHP} = \langle f\circ \phi(0) \rangle_{{\cal W_\alpha}} \, .
\ee
In the last equality we have used the definition
of the surface ${\cal W_\alpha}$, the semi-infinite cylinder of circumference $\alpha +1$,
\be
{\cal W}_\alpha \equiv \{ z   \, | \, {\rm Im} \, z \geq 0\,, \; z \sim z + \alpha +1 \, \} \,.
\ee
We also use the notation ${\cal C}_\ell$ for the semi-infinite cylinder of
circumference $\ell$,
\be \label{Cl}
{\cal C}_\ell \equiv \{ z   \, | \, {\rm Im} \, z \geq 0\,, \; z \sim z + \ell \, \} \,.
\ee
Clearly $ {\cal W}_{\alpha} = {\cal C}_{\alpha +1 }$. The  state $W_0 \equiv {\cal I}$
coincides with the identity of the star algebra; the  state $W_1 = |0 \rangle$
coincides with the SL(2) vacuum. The wedge states obey the abelian
algebra
\be
W_\alpha * W_\beta = W_{\alpha + \beta} \, .
\ee
In the operator formalism, the wedge states can be written as
\be
|W_\alpha \rangle = e^{-\frac{\alpha}{2}L^+} | {\cal I} \rangle =   e^{-\alpha L^+_L} | {\cal I} \rangle \, .
\ee
The derivations $L^+_L - L^+_R$ and $B^+_L - B^+_R$ annihilate the identity and
 commute with $L^+$, hence they annihilate all wedge states,
\be
(L^+_L - L^+_R) W_\alpha = (B^+_L - B^+_R) W_\alpha = 0 \,.
\ee 
Other important conservation laws are
\be
\left[  ( \alpha +1 )  L +   (\alpha-1) L^\star   \right]  \, W_\alpha = \left[  ( \alpha +1 )  B +   (\alpha-1) B^\star   \right]  \, W_\alpha = 0\,.
\ee
Some identities for the action of antighost operators on wedge states are:
\begin{eqnarray}
\label{bactiononwedges}
B W_\alpha   & =  & (1-\alpha) B^+_L W_\alpha\,,
 \\
 \label{bactiononwedges2}
(B - B^+_L) W_\alpha   & =   & -\alpha B^+_L W_\alpha \,,
\\ 
\label{bactiononwedges3}
B^\star W_\alpha  & =  & (1+\alpha) B^+_L W_\alpha\,,
\\
\label{bactiononwedges4}
(B^\star - B^+_L) W_\alpha  & =  & \alpha B^+_L W_\alpha  \,.
\end{eqnarray}
Some identities for the action of Virasoro exponentials on wedge states are:
\begin{eqnarray}
e^{- T L} W_\alpha   & =  & W_{e^{-T} (\alpha-1) +1}\,, \\
e^{- T L^\star} W_\alpha   & =  &  W_{e^T(\alpha +1)-1}\,, \\ 
e^{- T (L-L^+_L)} W_\alpha   & =   & W_{e^{-T} \alpha}\,, \\
e^{- T (L^\star-L^+_L)} W_\alpha    & =  &  W_{e^{T} \alpha} \,.
\end{eqnarray}

\section{Correlators on the cylinder}

We collect here some basic formulas for correlators on ${\cal C}_\ell$,
the semi-infinite cylinder of circumference $\ell$ defined in (\ref{Cl}). 
See also~\cite{yuji}. We introduce the notation  
\be
s_{ij} \equiv   \frac{ \sin ( \gamma  r_{ij} )}{\gamma}  \, , \quad
\gamma    \equiv  \frac{\pi}{\ell} \,    .
\ee
The basic ghost correlator is
\be
\label{ccc}
\bigl \langle c(r_1)\,  c(r_2) \, c(r_3) \, \bigr\rangle_{C_\ell} =    s_{12} \,  s_{13}\, s_{23} \, .
\ee
From this we readily find
\begin{eqnarray}
\bigl \langle c(r_1)  c(r_2)\partial c(r_3) \bigr\rangle_{C_\ell}   & =  &    - 
\frac{s_{12}  }{\gamma} \,  \sin(\gamma(r_1 + r_2 - 2 r_3))  \, ,\\
\bigl \langle c(r_1)  \partial c c(r_2) \bigr\rangle_{C_\ell}   & =  &    s_{12}^2 \,. 
\end{eqnarray}
The operator $(-B^+_L)$ is represented in the CFT language by an insertion of
${\cal B}$ \cite{yuji} defined by
\be
{\cal B } = \int_{V}  \frac{dz}{2 \pi i }\, b(z) \, ,  
\ee
where $V$  is a vertical line oriented downwards.
The operator ${\cal B}$ can be moved across a ghost insertion
with the help of the commutation relations
\be
{\cal B} c(r) = 1 - c(r) {\cal B} \, , \quad {\cal B} \partial c c(r) = - \partial c(r) + \partial c c(r) {\cal B} \,.
\ee
A ${\cal B}$ insertion at the beginning of a correlator  can be processed
as follows \cite{yuji}:
\be
\Bigl \langle {\cal B} \ldots  \ldots \Bigr \rangle_{{\cal C}_\ell} = - \frac{1}{\ell} \Bigl \langle \oint  \frac{dz}{2 \pi i} \, z b(z)
 \ldots \ldots \Bigr \rangle_{{\cal C}_\ell} \, ,
\ee 
where the contour is over the right boundary upwards
and the left boundary downwards.  Since $z b(z)$ transforms as a one-form,
this contour can be deformed into a sum of small contours, each 
 encircling a puncture with a ghost insertion. Finally, each contour integral is 
 evaluated with the help of the operator product expansion,
 \be
\oint \frac{dz}{2 \pi i} \, z b(z)  \, c(r) =  r \, , \quad  \oint \frac{dz}{2 \pi i} \, z b(z)  \, \partial c(r) =  1 \, , \quad
\oint \frac{dz}{2 \pi i} \, z b(z)  \, \partial c c(r) = c(r) - r \partial c (r)\,  ,
\ee
where in each case the contour is around the puncture at $z=r$, counterclockwise.
The simplest application of these techniques is to the correlator
\be \label{Bccc}
\langle {\cal B} \, c(r_1)  \, c(r_2) \, c(r_3) \, c(r_4) \rangle_{{\cal C}_\ell } = 
-\frac{1}{\ell} ( r_1 \, s_{23} s_{24} s_{34} - r_2 \, s_{13} s_{14} s_{24} + r_3 \, s_{12} s_{14} s_{24}  - r_4 \,s_{12} s_{13} s_{23} ) \, .
\ee

The matter correlator  for $n$ tachyon vertex operators is
\be  
\Bigl
\langle e^{i p_1\cdot X (r_1)}  
\ldots  e^{i p_n\cdot X (r_n)} \Bigr\rangle_{C_\ell} 
= (2\pi)^D  \delta \Bigl( \sum_{i=1}^n p_i\Bigr)
\prod_{i< j}  \bigl|   s_{ij}  \bigr|^{2\alpha' p_i\cdot p_j} \,.
\ee
For $n=4$, we can write  
\begin{eqnarray} \label{matter4pt}
\Bigl \langle e^{i p_1\cdot X (r_1)}  \ldots  e^{i p_4\cdot X (r_4)} \Bigr\rangle_{C_\ell} 
\hskip-10pt && =  (2\pi)^D   \delta \Bigl( \sum_{i=1}^4 p_i\Bigr) \cdot | s_{12}  |^{- \alpha'(p_1^2 + p_2^2 + \s)}
   | s_{13}  |^{\alpha'(\s +  \t + p_2^2 + p_4^2 )}  \\
 &&    \cdot  | s_{24}  |^{\alpha'(\s +  \t + p_1^2 + p_3^2 )}   
      | s_{23}  |^{-\alpha'( \t + p_2^2 + p_3^2 )}         | s_{14}  |^{-\alpha'( \t + p_1^2 + p_4^2 )} 
  |  s_{34}  |^{- \alpha'(p_3^2 + p_4^2 + \s)} \, , \nonumber
\end{eqnarray}
where $\s$, $\t$, $\u$ are the familiar Mandelstam invariants
\be
\s = - (p_1+p_2)^2  \,, ~ \t = - (p_2+ p_3)^2  \,, ~ \u = - (p_1 + p_3)^2 \,, \quad 
\s+ \t + \u = - \sum p_i^2 \,.
\ee

\section{Proof of the rearrangement formula}

By the CBH formula, we can reorder the exponentials in (\ref{blbsls}) as
\be 
 e^{-T_1 L} \, e^{-T_2 L^\star} =  e^{-\tau_2 L^\star} \, e^{-\tau_1 L}  \, ,
\ee
where the parameters $\tau_i$ are appropriate functions of the parameters $T_i$.
A simple way to do the calculation is to represent $L$ and $L^\star$  in the adjoint
representation, using (\ref{xyadjoint}) with $L=x$ and $ L^\star = y-x$.
It is convenient to introduce the notation
\be
\bar t_i  \equiv e^{-T_i}  \, ,\quad  t_i \equiv  e^{-\tau_i} \, , \qquad i = 1,2 \,.
\ee
We find
\be
t_i   =     \frac{\bar t_i}{\bar t_1 + \bar t_2 - \bar t_1 \bar t_2 } \,.
\ee
The inverse relations are
\be
\bar t_1 = \frac{t_1 + t_2 -1}{t_2} \, ,\quad \bar t_2 = \frac{t_1 + t_2 -1}{t_1} \, .
\ee
By another application of CBH we find
\be
\label{belstar}
B e^{- \tau L^\star} = e^{-\tau L^\star}  \left( e^{-\tau} B + (e^{-\tau} -1) B^\star \right) \, ,
\ee
and the BPZ conjugate equation
\be
 e^{- \tau L} B^\star =   \left(  e^{-\tau} B^\star + (e^{-\tau} -1) B \right) e^{-\tau L} \, .
\ee
Back in (\ref{blbsls}),
\begin{eqnarray}
\frac{B}{L}  \frac{B^\star}{L^\star} = \int_0^\infty dT_1 dT_2 \; (1- t_1 - t_2) \, e^{-\tau_2 L^\star} \, B^\star \; B   \, e^{-\tau_1 L} \,.
\end{eqnarray}
The integration measure can be written as
\be
dT_1 dT_2 (1-t_1 - t_2) = \frac{ d \bar t_1 d \bar t_2}{\bar t_1 \bar t_2}  (1-t_1 -t_2) =- \frac{dt_1 dt_2}{t_1 t_2} =- d\tau_1 d\tau_2 \,.
\ee
Here we have used the Jacobian determinant
\be
{\rm det} \frac{\partial (\bar t_1, \bar t_2) }{\partial ( t_1, t_2 ) } = \frac{t_1 + t_2 -1}{t_1^2 t_2^2} = \frac{\bar t_1 \bar t_2}{(t_1 t_2)(t_1+t_2 -1)}\,.
\ee
Note that in the new variables the integration region  is given by
\be
{\cal M}   \equiv \{ \tau_1 , \tau_2  \; |  \tau_1 \geq 0\,, \tau_2 \geq 0\,, e^{-\tau_1} + e^{-\tau_2} \geq 1 \} \,.
\ee
All in all, we find the simple result
\be 
\frac{B}{L}  \frac{B^\star}{L^\star} =- \int_{{\cal M} } d \tau_1 d \tau_2 \;   B^\star  e^{-\tau_2 L^\star} \; B   \, e^{-\tau_1 L} \,.
\ee

\end{document}